\documentclass[twocolumn,prb,showpacs,floatfix]{revtex4}
\usepackage{graphicx}
\usepackage{amssymb,amsmath,latexsym}
\usepackage{bm}
\usepackage[dvips]{epsfig}

\begin{document}

\title{Ballistic spin field-effect transistors: Multichannel effects}
\author{Jae-Seung Jeong}
\email{jsjeong@postech.ac.kr}
\author{Hyun-Woo Lee}
\affiliation{PCTP and Department of Physics, Pohang University of Science
and Technology, Pohang, Kyungbuk 790-784, Korea}

\begin{abstract}
We study a ballistic spin field-effect transistor (SFET) with
special attention to the issue of multichannel effects. The
conductance modulation of the SFET as a function of the Rashba
spin-orbit coupling strength is numerically examined for the
number of channels ranging from a few to close to 100.
Even with the ideal spin injector and collector,
the conductance modulation ratio, defined as
the ratio between the maximum and minimum conductances, decays rapidly and approaches one
with the increase of the channel number.
It turns out that the decay is considerably faster
when the Rashba spin-orbit coupling is larger.
Effects of the electronic coherence are also examined in the multichannel regime
and it is found that the coherent Fabry-Perot-like interference in the multichannel regime
gives rise to a nested peak structure.
For a nonideal spin injector/collector structure,
which consists of a conventional metallic ferromagnet-thin insulator-2DEG heterostructure,
the Rashba-coupling-induced conductance modulation is strongly affected
by large resonance peaks that arise from the electron confinement effect of the insulators.
Finally scattering effects are briefly addressed and it is found that
in the weakly diffusive regime, the positions of the resonance peaks fluctuate,
making the conductance modulation signal sample-dependent.

\end{abstract}

\maketitle

\section{Introduction}
\label{Introduction}
At an interface of a semiconductor heterostructure,
electrons confined to the interface form a two-dimensional electron gas (2DEG).
The Rashba spin-orbit (RSO) coupling~\cite{Bychkov84JPC} arises in the 2DEG
when the interface confinement potential breaks the structural inversion symmetry.
It is demonstrated that the strength of the RSO coupling can be modulated~\cite{Nitta97PRL}
by a gate voltage that affects the asymmetry of the confinement potential.
The RSO coupling can be a useful tool for spintronic applications in semiconductors~\cite{Zutic04RMP}.
One of representative examples is the spin field-effect transistor (SFET)
proposed by Datta and Das~\cite{Datta90APL},
which is based on the spin precession by the RSO coupling within the 2DEG and
on the spin preparation and detection by two spin-selective electrodes (injector and collector).

Despite intensive experimental efforts, the SFET has not been realized yet
and there are numerous reports addressing practical problems for the realization of the SFET
such as the spin injection/detection~\cite{Filip00PRB,Schmidt00PRB}
and the spin relaxation~\cite{Bournel97SSC,Kiselev00PRB,Cahay04PRB}.
In this paper, we 
focus on the issue of the multichannel effects.
While a narrow 2DEG with only one transport channel is an ideal environment
for the SFET operation~\cite{Bandyopadhyay04APL},
a single-channel system is rather difficult to realize in experiments.
For a 2DEG with the Fermi wavelength of 100 A, for instance,
the width of the system needs to be smaller than 100 A for the system to be in the single channel regime.
Preparing a 2DEG with its width less than 100 A is, though not impossible, quite demanding.
The experimental difficulty is further enhanced when it is taken into account that
for the transistor operation,
the system needs to be longer than a certain minimum length
for the spin precession angle to be of order $\pi$ at least.
For the 2DEG formed at the InGaAs/InAlAs interface, for instance,
the minimum length was estimated to  be 0.67 $\mu$m~\cite{Datta90APL}.
Taking into account the experimental difficulty in the preparation of
a sufficiently long single-channel system, a multichannel system is thus a more practical test ground
for the SFET.

Behaviors of a multichannel SFET may be crucially affected
by the inter-channel coupling.
In Ref.~\cite{Datta90APL}, it was argued that effects of the inter-channel coupling will be
negligible and a multichannel SFET will behave as a collection of uncoupled single-channel SFETs
when a dimensionless number $2m^*\alpha w/\hbar^2$ is sufficiently smaller than
$2\pi$, where $w$ is the width of the 2DEG, $m^*$ is the effective mass of the electron in the 2DEG,
and  $\alpha$ is the RSO coupling parameter.
In this paper, we examine systematically behaviors of a multichannel SFET
with the number of channels from a few to close to 100
and with $\alpha$ from $10^{-12}$ eVm to $10^{-10}$ eVm,
covering most reported values of $\alpha$~\cite{Luo90PRB,Engels97PRB,Heida98PRB,Schultz96SST,Nitta97PRL}.
For the examined range, the value of $2m^*\alpha w/\hbar^2$ varies
from approximately 0.03 to 50,
including the range where the inter-channel coupling is not negligible.
Our result thus provides a systematic investigation of
the multichannel effects.
We also examine effects of the electronic coherence in the multichannel regime.
For a single-channel SFET, it was recently demonstrated~\cite{Schapers01PRB,Mireles02EL,Larsen02PRB,Lee05PRB}
that the electronic coherence makes the conductance
deviate from the sinusoidal dependence on $\alpha$.
We find that the electronic coherence in the multichannel regime can give rise to
a nested peak structure.
Our analysis indicates that the nested peak structure is
due to the coherent Fabry-Perot-like interference
in the multichannel regime.
We remark that the nested peak structure was not found in previous studies~\cite{Bournel97SSC,Kiselev00PRB,Mireles01PRB} of a multichannel SFET
since effects of the electronic coherence were not properly taken into account.


The paper is organized as follows. Section~\ref{conductance} reports
the numerical conductance calculation results of the conductance of multichannel SFETs
as a function of the RSO coupling parameter $\alpha$.
SFETs with various channel numbers are examined.
When the spin injector and collector in SFETs are ideal,
it is found that the conductance modulation ratio between the maximum and minimum conductances
reduces as the number of channel increases.
The decaying rate of the ratio is significantly larger for SFETs with larger $\alpha$.
In order to gain insights into the numerical calculation results,
energy dispersion relations and spin configurations of energy eigenstates are examined.
Section~\ref{CoherentSFET} focuses on the conductance properties of SFETs with rather small number of channels
that exhibit a large conductance modulation ratio.
It is found that the electronic coherence gives rise to a nested peak structure.
Effects of an in-plane magnetic field are also addressed briefly
and it is verified that the magnetic-field-induced peak splitting,
reported previously for a single-channel SFET~\cite{Lee05PRB},
persists in the multichannel regime.
Section~\ref{nonidealSFET} addresses briefly behaviors of multichannel SFETs equipped
with nonideal spin injectors and collectors, which consist of
conventional metallic ferromagnet-thin insulator-2DEG hybrid structures.
Scattering effects are also addressed briefly.
Section~\ref{Summary} summarizes the paper.

\section{Conductance of multichannel SFET}
\label{conductance}
\begin{figure}[b!]
\includegraphics[width=6cm]{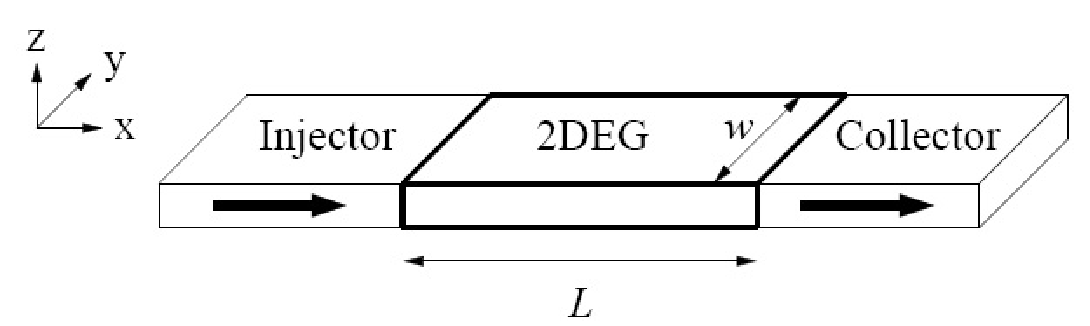}
\caption{
Schematic diagram of a ballistic spin field effect transistor
consisting of two-dimensional electron gas (2DEG) of finite width $w$
and two spin-selective electrodes (injector and collector),
whose spins are assumed to be fully polarized
along the $(+x)$-direction.
The strength of the Rashba coupling can be tuned by external gates (not shown).
}
\label{SFET}
\end{figure}
Figure~\ref{SFET} shows a schematic drawing of a SFET with finite width $w$.
Within the 2DEG in the $xy$-plane, the effective Hamiltonian of electrons reads
\begin{equation} \label{eq:h1}
H_{\rm 2D}=\frac{\mathbf{p}^2}{2m^*}\!+\!\frac{\alpha}{\hbar}
(\vec{\sigma}\times \mathbf{p})\cdot\mathbf{\hat{z}}\!+\!V_{\rm c}(y),
\end{equation}
where $\textbf{p}\!=\!(p_x, p_y)$ is the momentum operator within
the 2DEG, $V_{\rm c}(y)$ is the transverse confinement potential
with $V_{\rm c}(y)=0$ for $0<y<w$ and $V_{\rm c}(y)=\infty$
otherwise.
Here
$\vec{\sigma}=(\sigma_x, \sigma_y, \sigma_z)$ is the Pauli spin operator,
and $\mathbf{\hat{z}}$ a unit vector perpendicular to the 2DEG.

In order to focus on the issue of the multichannel transport,
we assume situations where effects of other practical problems are minimized;
scattering by impurities and phonons is ignored in $H_{\rm 2D}$
and ideal spin injection and detection are assumed.
To be specific, an electron in the injector and collector is assumed to be subject to
the following Hamiltonian,
\begin{equation}
H_{\rm inj/col}=\frac{\mathbf{p}^2}{2m^*}+V_{\rm c}(y)-g\mu_B B_{\rm ex}(\sigma_x-1),
\end{equation}
where $B_{\rm ex}$ is the effective exchange energy in the injector/collector,
$g$ is the Lande's g-factor, and $\mu_B$ is the Bohr magneton.
For sufficiently large $g\mu_B B_{\rm ex}$ ($>$0), electrons in the injector and collector
are 100\% spin-polarized along the $(+x)$-direction.
Note that with this choice of $H_{\rm inj/col}$,
the Fermi wavelength of $(+x)$-polarized electron within the injector/collector
matches perfectly with that within the 2DEG when $\alpha=0$ and thus
the so-called conductance mismatch problem for the spin injection/detection does not appear for the 2DEG-injector(collector) interface.
For nonzero $\alpha$, there is a weak mismatch in the Fermi wavelength but
for the range of $\alpha$ values examined below ($\alpha<10^{-10}$ eVm), the conductance mismatch problem
turns out to be a weak effect (see Appendix~\ref{IdealInjCol} for details).

\subsection{Numerical conductance calculation}
\label{Transport}
For the numerical conductance calculation
of the multichannel SFETs,
we use the tight-binding (TB) Hamiltonian~\cite{Datta95Book,Pareek02PRB}
$H^{\rm TB}=H_{\rm 2D}^{\rm TB}+H_{\rm inj}^{\rm TB}+H_{\rm col}^{\rm TB}+H_{\rm coupling}^{\rm TB}$,
where $H_{\rm 2D}^{\rm TB}$ amounts to the TB approximation
of the Hamiltonian $H_{\rm 2D}$ [Eq.~(\ref{eq:h1})] for the 2DEG,
\begin{eqnarray}
\label{eq:2DTB}
H_{\rm 2D}^{\rm TB}&=&
4t\sum_{i=1}^{N_x}\sum_{j=1}^{N_y}\sum_{s_z=\pm 1}c_{i,j,s_z}^\dagger c_{i,j,s_z} \\
& &-t \sum_{s_z=\pm 1}
\left[ \sum_{i=1}^{N_x-1}\sum_{j=1}^{N_y}
\left( c_{i+1,j,s_z}^\dagger c_{i,j,s_z}
  +{\rm H.c.} \right)\right.\nonumber \\
& & +\left.\sum_{i=1}^{N_x}\sum_{j=1}^{N_y-1}
\left(c_{i,j+1,s_z}^\dagger c_{i,j,s_z}+ {\rm H.c.} \right) \right] \nonumber \\
& &+\lambda
\sum_{s_z=\pm 1}\sum_{s'_z=\pm 1}
\left\{-\sum_{i=1}^{N_x-1}\sum_{j=1}^{N_y}
\left[c_{i+1,j,s_z}^\dagger (i\sigma_y)^{s_z,s'_z}
  c_{i,j,s'_z} \right.\right.\nonumber\\
& &\left.\left.+{\rm H.c.} \right] \right. \nonumber \\
& & \left. + \sum_{i=1}^{N_x}\sum_{j=1}^{N_y-1} \left[
 - c_{i,j+1,s_z}^\dagger (i\sigma_x)^{s_z,s'_z}
c_{i,j,s'_z}+{\rm H.c.}\right]\right\}, \nonumber
\end{eqnarray}
where $c_{i,j,s_z}$ is the annihilation operator
of an electron at ${\bf r}_{ij}=a(i\hat{\bf x}+j\hat{\bf y})$ with spin $s_z$ along the $z$-axis,
$a$ is the lattice spacing used for the TB approximation,
and $N_x$ and $N_y$ are related to the length $L$ and width $w$
of the 2DEG via $a(N_x+1)=L$
and $a(N_y+1)=w$, respectively.
With the choice $t=\hbar^2/2m^*a^2$ and $\lambda=\alpha/2a$,
Eq.~(\ref{eq:2DTB}) reduces to Eq.~(\ref{eq:h1}) in the limit $a\rightarrow 0$.
The TB Hamiltonians for the injector and collector are similarly given by
\begin{eqnarray}
\label{eq:InjTB}
H_{\rm inj}^{\rm TB} &=&
4t \sum_{i=0}^{-\infty}\sum_{j=1}^{N_y}c_{i,j,s_x=1}^\dagger c_{i,j,s_x=1} \\
& &-t
\left[ \sum_{i=-1}^{-\infty}\sum_{j=1}^{N_y}
\left( c_{i+1,j,s_x=1}^\dagger c_{i,j,s_x=1}
  +{\rm H.c.} \right)\right. \nonumber\\
& &\left.+\sum_{i=0}^{-\infty}\sum_{j=1}^{N_y-1}
\left(c_{i,j+1,s_x=1}^\dagger c_{i,j,s_x=1}+ {\rm H.c.}\right) \right], \nonumber
\end{eqnarray}
\begin{eqnarray}
\label{eq:ColTB}
H_{\rm col}^{\rm TB} &=&
4t \sum_{i=N_x+1}^{\infty}\sum_{j=1}^{N_y}c_{i,j,s_x=1}^\dagger c_{i,j,s_x=1} \\
& &-t
\left[ \sum_{i=N_x+1}^{\infty}\sum_{j=1}^{N_y}
\left( c_{i+1,j,s_x=1}^\dagger c_{i,j,s_x=1}
  +{\rm H.c.} \right)\right. \nonumber\\
& &\left.+\sum_{i=N_x+1}^{\infty}\sum_{j=1}^{N_y-1}
\left(c_{i,j+1,s_x=1}^\dagger c_{i,j,s_x=1}+ {\rm H.c.}\right) \right], \nonumber
\end{eqnarray}
where $c_{i,j,s_x=\pm 1}\equiv (c_{i,j,s_z=1}\pm c_{i,j,s_z=-1})/\sqrt{2}$.
Note that the electron operators for $s_x=-1$ do not appear in $H_{\rm inj/col}^{\rm TB}$,
since the injector/collector is assumed to be 100\% spin polarized along $(+x)$-direction.
The coupling Hamiltonian between the 2DEG and the injector/collector is given by
\begin{eqnarray}
\label{eq:couplingTB}
H_{\rm coupling}^{\rm TB}&=&
-t \sum_{j=1}^{N_y}
\left( c_{1,j,s_x=1}^\dagger c_{0,j,s_x=1}\right.\nonumber\\
& &\left.+c_{N_x+1,j,s_x=1}^\dagger c_{N_x,j,s_x=1} + {\rm H.c.} \right).
\end{eqnarray}
Note that electrons whose spin is pointing along the $(-x)$-direction are not allowed
to hop between the 2DEG and the injector(collector).
Note also that the hopping parameter $t$ in $H_{\rm inj/col}^{\rm TB}$
is the same as that in $H_{\rm 2D}^{\rm TB}$.
Thus when effects of $\alpha$ (or $\lambda$) on the Fermi wavelength is not significant,
the conductance mismatch problem should be minimal (see Appendix~\ref{IdealInjCol} for detail),
allowing ideal spin injection and detection.

The Landauer-B\"uttiker formalism is used for the numerical conductance calculation
and the method in Ref.~\cite{Anantram98PRB} is used to evaluate matrix elements
of related Greens' functions.
The following parameters are used;
$m^*=0.04 \times m_{\rm electron}$, where $m_{\rm electron}$ is the free electron mass,
and the Fermi energy $E_F=0.103$ eV, which, in the absence of $\alpha$ and $V_{\rm c}$,
amounts to the Fermi wavelength $2\pi/k_F=191$ A and
the electron density
of $n_s=1.72\times 10^{12}$ cm$^{-2}$ in the 2DEG.
For the TB approximation, we choose $a=(2\pi/k_F)/9.95$.
Thus the hopping energy $t=\hbar^2/2m^*a^2$ becomes $0.259$ eV.

Below the conductance is first evaluated in two windows of $\alpha$,
(0$\sim$10)$\times 10^{-12}$ eVm and (45$\sim$55)$\times 10^{-12}$ eVm.
In various candidate systems for the SFET
such as InGaAs/InAlAs~\cite{Nitta97PRL}, InAs/GaSb~\cite{Luo90PRB},
InGaAs/InP~\cite{Engels97PRB}, and InAs/AlSb~\cite{Heida98PRB},
$\alpha$ typically falls within the first window.
A recent calculation~\cite{Silva03PRB} also reported that for III-V quantum wires,
$\alpha$ is typically in this weak coupling regime.
Search for systems with higher $\alpha$ is in progress
and effects of strong RSO coupling on the energy spectrum
are addressed theoretically~\cite{Moroz99PRB}.
For CdTe/HgTe/CdTe, for instance,  values up to $\sim 45\times 10^{-12}$ eVm has been reported~\cite{Schultz96SST},
which motivates the second window.
For a SFET with relatively short $L$, the conductance is
evaluated in a much wider window of $\alpha$,
(0$\sim$100)$\times 10^{-12}$ eVm.

\begin{figure}[b]
\centerline{\includegraphics[width=9cm]{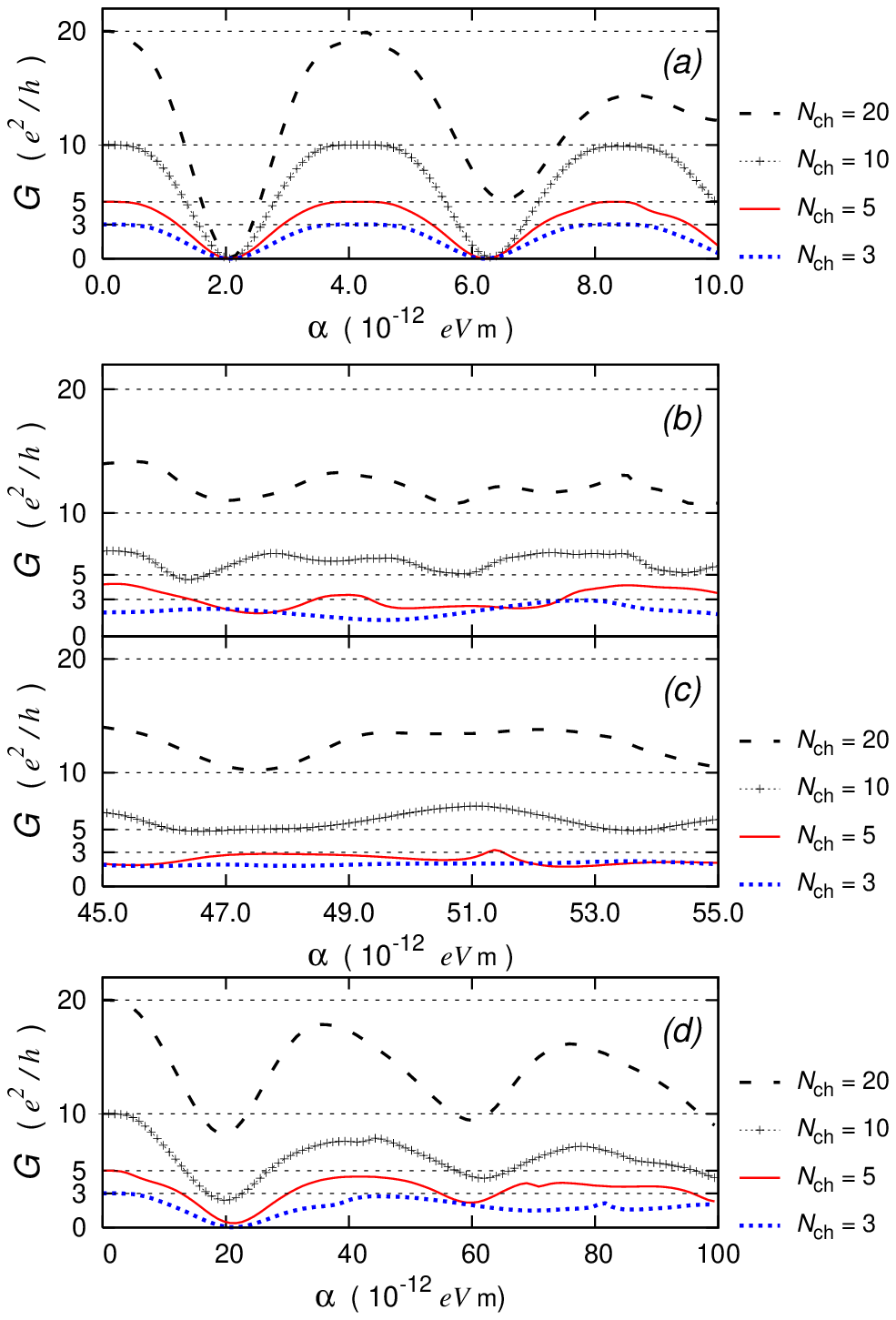}}
\caption{(Color online) Conductance $G$ of a ballistic multichannel SFET
as a function of the RSO coupling parameter $\alpha$ at the temperature $T$=0.
Ideal spin injection and detection are assumed.
The following
parameters are used; the effective mass $m^*=0.04\times m_{\rm electron}$,
where $m_{\rm electron}$ is the free electron mass, and
the electron density
$n_s=1.72\times 10^{12}$ cm$^{-2}$.
The length $L$ of the 2DEG is $1.44\,\mu$m for (a) and (b),
$0.8\, \mu$m ($\approx 0.556\times 1.44\,\mu$m) for (c), and
$0.1\times 1.44\,\mu$m for (d).
}
\label{multiG01}
\end{figure}

\begin{figure}[t]
\centerline{\includegraphics[width=8cm]{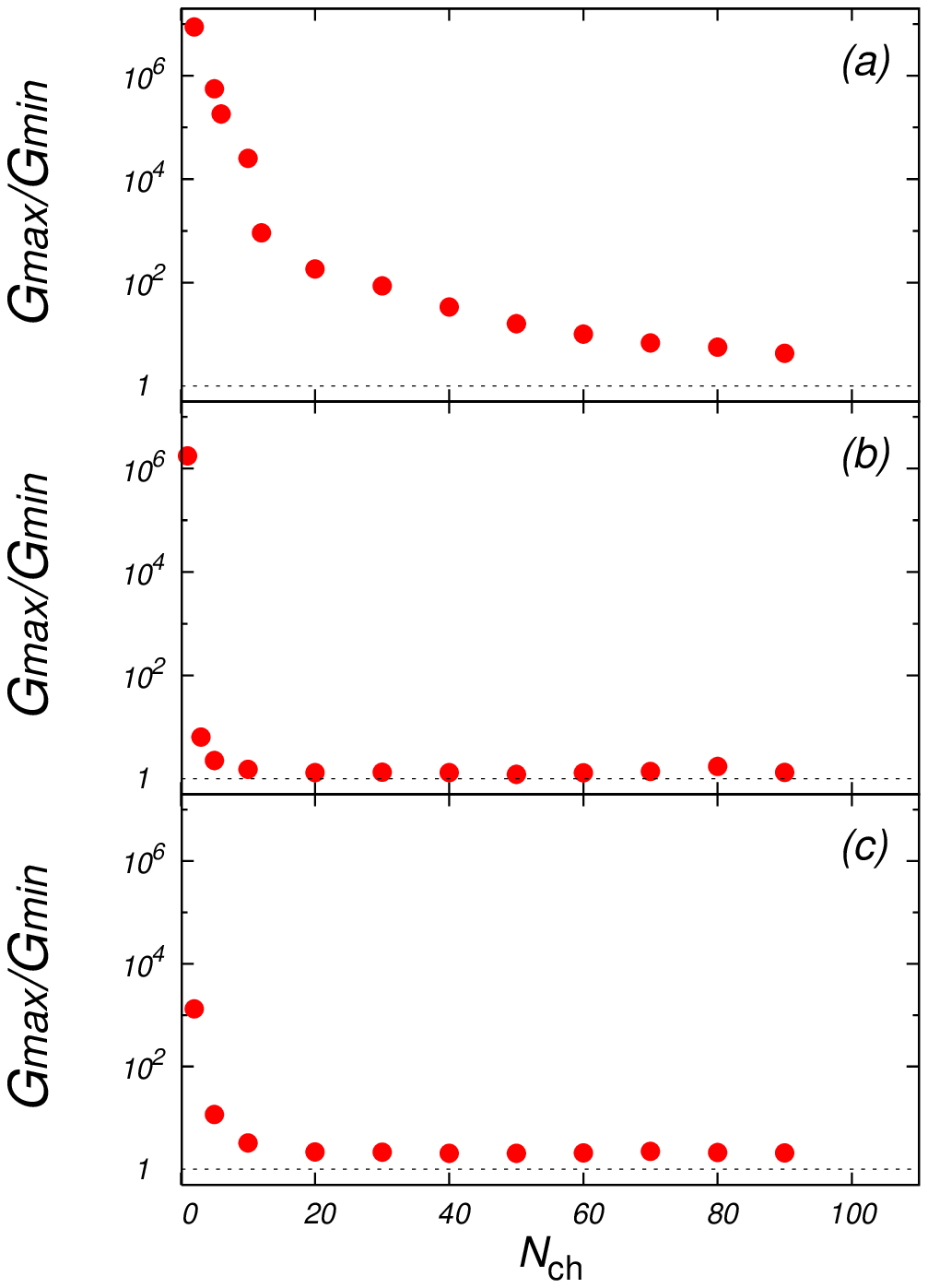}}
\caption{(Color online) The ratio between the maximum conductance $G_{\rm max}$
and the minimum conductance $G_{\rm min}$
as a function of the number of channels $N_{\rm ch}$.
The panels (a), (b), and (c) correspond to the situations
in Fig.~\ref{multiG01}(a), (b), and (d), respectively.
The parameters are the same as those for Fig.~\ref{multiG01}.
}
\label{multiG02}
\end{figure}

Figure~\ref{multiG01} shows the evolution of the conductance $G$ as a function of $\alpha$
for the SFET with 3, 5, 10, and 20 channels, respectively.
For the range of $\alpha$ examined below,
the number of channels $N_{\rm ch}$ is related to the width $w$ of the 2DEG
via the relation~\cite{Datta95Book}
$N_{\rm ch}={\rm Int}(k_F w/\pi)$,
where ${\rm Int}(x)$ denotes the largest integer not exceeding $x$.
The length $L$ of the 2DEG is assumed to be
$1.44\, \mu$m for (a) and (b), and $0.8 \, \mu$m ($\approx 0.556 \times 1.44 \, \mu$m)
for (c), and $0.1 \times 1.44\, \mu$m for (d), respectively.
In Fig.~\ref{multiG01}(a), the SFET is operated in
the $\alpha$ range from 0 to $10\times 10^{-12}$ eVm.
To estimate the spin precession angle,
we use the formula $2m^* \alpha L/\hbar^2$.
Though this formula is derived in the single-channel limit,
it may still be useful to estimate the spin precession in multichannel SFETs.
Then the average spin precession angle,
estimated from the formula $2m^*\alpha L/\hbar^2$ with the average $\alpha=5.0 \times 10^{-12}$ eVm,
is around $1.2\times 2\pi$, and the variation of the spin precession angle
within the specified range is $\pm 1.2 \times 2\pi$.
Note that the SFET exhibits an almost ideal conductance modulation
in the sense that
the maximum of $G$ reaches almost $(e^2/h)N_{\rm ch}$
and the minimum of $G$ reaches almost zero.
The conductance modulation behavior becomes, however, less ideal
with the increase of $N_{\rm ch}$.
For $N_{\rm ch}=20$, for instance, Fig.~\ref{multiG01}(a) already shows some deviation from the ideal behavior.
Figure~\ref{multiG02}(a) shows, as a function of $N_{\rm ch}$,
the ratio between $G_{\rm max}$ and $G_{\rm min}$,
where $G_{\rm max}$ and $G_{\rm min}$ are the maximum and minimum values of $G$
in the interval $1.8\times 10^{-12}$ eVm $< \alpha <$ $4.5 \times 10^{-12}$ eVm
in Fig.~\ref{multiG01}(a).
It clearly shows the decay of the ratio with $N_{\rm ch}$.

In Fig.~\ref{multiG01}(b) (again with $L=1.44$ $\mu$m),
the SFET is operated in the $\alpha$ range
$(45\sim 55)\times 10^{-12}$ eVm.
Note that the width of the $\alpha$ range is the same as that in Fig.~\ref{multiG01}(a).
Since $\alpha$ is larger than the values in Fig.~\ref{multiG01}(a),
the spin precession angle will be larger.
The average spin precession angle in this range,
estimated from the formula $2m^*\alpha L/\hbar^2$ with the average $\alpha=50\times 10^{-12}$ eVm,
is around $12\times 2\pi$, about 10 times larger than in Fig.~\ref{multiG01}(a).
On the other hand, the variation of the spin precession angle
within the specified range is about $\pm 1.2 \times 2\pi$,
same as in Fig.~\ref{multiG01}(a).
Note that
the conductance modulation behavior is now much less ideal;
the maximum and minimum of $G$ deviate considerably
from $(e^2/h)N_{\rm ch}$ and $0$, respectively.
Moreover the conductance oscillation with $\alpha$ becomes
somewhat irregular.
Figures~\ref{multiG02}(b) shows, as a function of $N_{\rm ch}$,
the ratio between $G_{\rm max}$ and $G_{\rm min}$ for the SFET in Fig.~\ref{multiG01}(b). 
Here $G_{\rm max}$ and $G_{\rm min}$ are evaluated
in the $\alpha$ range $(48.5\sim 51.5)\times 10^{-12}$ eVm. 
Note that the decay of the ratio with increasing $N_{\rm ch}$ is much faster
than that in Fig.~\ref{multiG02}(a).
For small $N_{\rm ch}$ around 5,
the ratio $G_{\rm max}/G_{\rm min}$ already drops to of order unity.

One intuitive explanation for the much less ideal behavior
in Figs.~\ref{multiG01}(b) and \ref{multiG02}(b),
compared to Figs.~\ref{multiG01}(a) and \ref{multiG02}(a),
is fluctuations of the spin precession rates from channel to channel.
For example, if the spin precession angles per length deviate from $2m^*\alpha/\hbar^2$
and the deviations fluctuate from channel to channel,
the conductance modulation may become less ideal.
To test this possibility, we examine a SFET with shorter $L$
since effects of the channel-to-channel fluctuation are expected to be weakened
with decreasing $L$.
In Fig.~\ref{multiG01}(c),
the length $L$ of the 2DEG is reduced to $0.8 \, \mu$m ($\approx 0.556 \times 1.44\,\mu$m).
When the SFET is operated in the same $\alpha$ range as in Fig.~\ref{multiG01}(b),
the average spin precession
is estimated to $6.7\times 2\pi$,
which is about 44\% smaller than that in Fig.~\ref{multiG01}(b),
and the variation of the spin precession angle within the specified range is estimated to $\pm 0.7\times 2\pi$.
Figure~\ref{multiG01}(c) shows that the conductance modulation is still far from ideal
despite the 44\% reduction in the average spin precession angle.

In Fig.~\ref{multiG01}(d),
the length $L$ of the 2DEG is further reduced to $0.1\times 1.44\,\mu$m,
one tenth of that in Fig.~\ref{multiG01}(a).
When the SFET is operated in
the range $0<\alpha<100\times 10^{-12}$ eVm, which is 10 times larger than in Fig.~\ref{multiG01}(a),
the average spin precession angle and the variation of the angle within the specified $\alpha$ range
are exactly same as those in Fig.~\ref{multiG01}(a).
Thus
the situations in Figs.~\ref{multiG01}(a) and (d) are identical
as far as the expression $2m^*\alpha L/\hbar^2$ for the spin precession angle is concerned.
The numerical calculation result in Fig.~\ref{multiG01}(d) indicates however that
while the conductance modulation behavior is somewhat improved compared to those in Figs.~\ref{multiG01}(b) and (c),
it is still less ideal than that in Fig.~\ref{multiG01}(a).
Figure~\ref{multiG02}(c) shows the $N_{\rm ch}$-dependence of the ratio $G_{\rm max}/G_{\rm min}$
for the situation in Fig.~\ref{multiG01}(d),
where $G_{\rm max}$ and $G_{\rm min}$ are evaluated
in the interval $18\times 10^{-12}$ eV $< \alpha <$ $45 \times 10^{-12}$ eVm.
Note that the decay of the ratio is
still much faster than in Fig.~\ref{multiG02}(a).
This result leads one to conclude
that the channel-to-channel fluctuations of the spin precession angles are
not the main reason for the deviations from the ideal behavior.

\subsection{Transport channels}
\label{TransportChannels}
\begin{figure}[b!]
\centerline{\includegraphics[width=8cm]{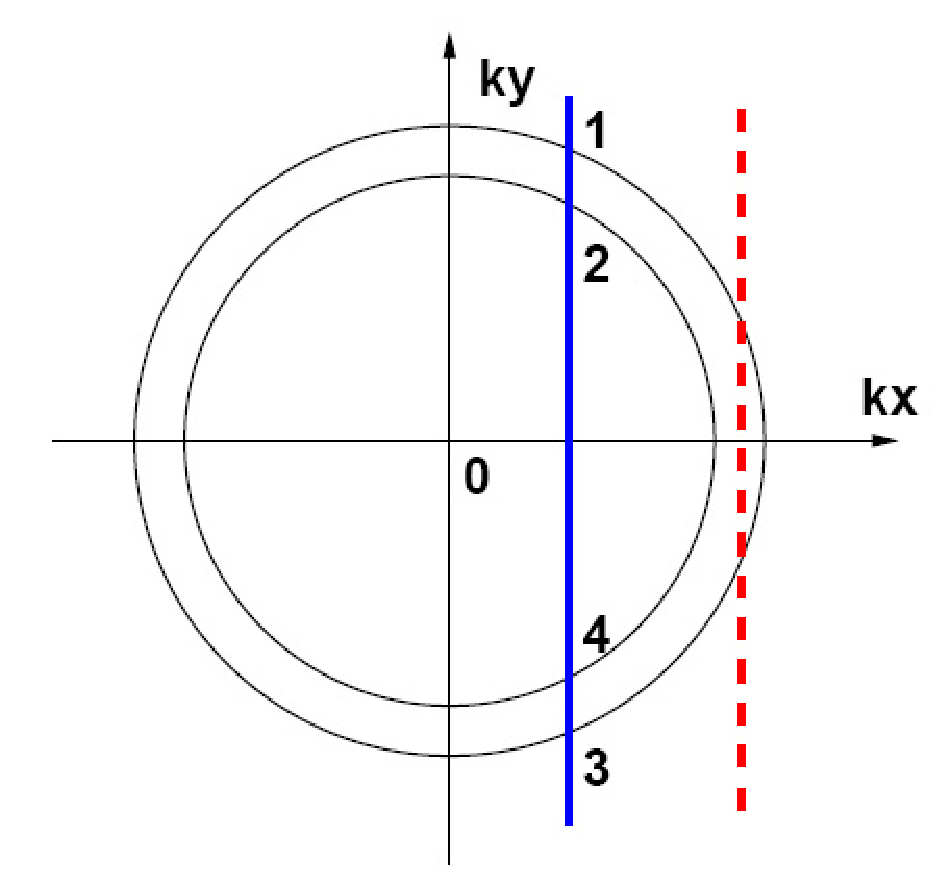}}
\caption{(Color online) Four solutions $k_{y,j}$ ($j=1,2,3,4$)
of $\hbar^2 (k_x^2+k_{y,j}^2)/2m^*+(-1)^j \alpha (k_x^2+k_{y,j}^2)^{1/2}=E$,
which is the energy dispersion relation
in the absence of $V_{\rm c}(y)$.
Note that for certain $k_x$ (denoted by the red dashed line),
two solutions become purely imaginary numbers.
}
\label{2D}
\end{figure}

In order to gain an insight into the numerical results in Figs.~\ref{multiG01} and \ref{multiG02},
it is useful to examine
energy eigenfunctions and eigenenergies of $H_{\rm 2D}$ [Eq.~(\ref{eq:h1})].
Let $\Psi(x,y)=e^{ik_x x} \psi(y)$ denote a eigen wavefunction with energy $E$.
Due to the translational symmetry along the $x$-direction within the 2DEG,
it can be expressed as a superposition of four plane waves
$\Psi=\sum_{j=1}^{4}c_j \Psi_{{\bf k}_j}$,
where the plane wave $\Psi_{{\bf k}_j}(x,y)=e^{i{\bf k}_j\cdot {\bf r}}u_j({\bf k}_j)$
is an eigenstate of $H_{\rm 2D}$ in the absence of $V_{\rm c}(y)$.
Here all four  wavevectors ${\bf k}_j=(k_{x,j},k_{y,j})$ share the same longitudinal component $k_{x,j}=k_x$
and their transverse components $k_{y,j}$ are determined from the  relation
$\hbar^2 (k_{x,j}^2+k_{x,j}^2)/2m^*+(-1)^j \alpha (k_{x,j}^2+k_{y,j}^2)^{1/2}=E$,
which is the energy-dispersion relation in the absence of $V_{\rm c}(y)$ (see Fig.~\ref{2D}).
The spinor $u_j ({\bf k})$ is given by
\begin{equation}
\label{eq:spinor02}
u_j ({\bf k})={1\over\sqrt{2}}
\left(
\begin{array}{c}
 1 \\ (-1)^{j+1}i e^{i\varphi({\bf k})}
\end{array}
\right),
\end{equation}
where $\varphi({\bf k})$ is defined by
$\cos{\varphi({\bf k})}=k_x/(k_{x}^2+k_{y}^2)^{1/2} $
and
$\sin{\varphi({\bf k})}=k_y/(k_{x}^2+k_{y}^2)^{1/2} $.
Then the four constraints from the boundary conditions $\Psi(x,0)=\Psi(x,w)=0$
fix the four coefficients $c_j$,
and the exact energy-dispersion relations and the eigen wavefunctions can be obtained.
This procedure can be simplified further by
using additional symmetries of $H_{\rm 2D}$.
See Appendix~\ref{Symmetry} for further details on the symmetries,
and Appendix~\ref{ExactDispersion} for the use of the symmetries for the evaluation
of the exact energy-dispersion relation and the eigen wavefunctions.

\begin{figure}[b!]
\centerline{\includegraphics[width=9cm]{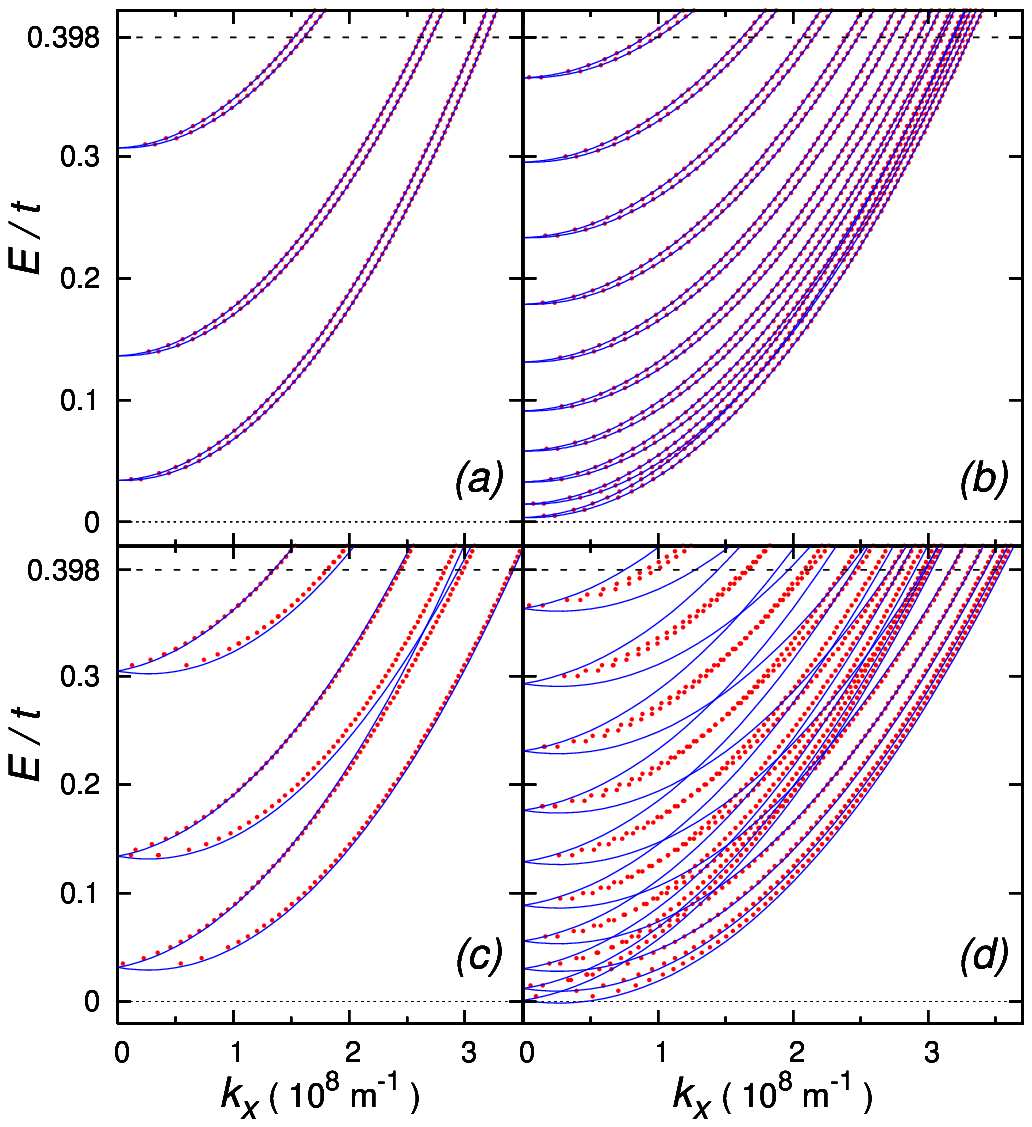}}
\caption{(Color online) The energy dispersion relations in a Rashba spin-orbit (RSO)
coupled system with hard wall confinement. Red dotted lines represent the exact
dispersion relations, while blue solid lines the perturbative results,
Eq.~(\ref{eq:eigenenergy}).
The two types of the lines overlap almost perfectly
in (a) and (b).
The RSO coupling parameter $\alpha$ and the number of channels $N_{\rm ch}$ are
$\alpha=8.3\times 10^{-12}$ eVm, $N_{\rm ch}=3$ in (a),
$\alpha =8.3\times 10^{-12}$ eVm, $N_{\rm ch}=10$ in (b),
$\alpha=49.8\times 10^{-12}$ eVm, $N_{\rm ch}=3$ in (c), and
$\alpha=49.8\times 10^{-12}$ eVm, $N_{\rm ch}=10$ in (d), respectively.
The width $w$ is $32.7$ nm for (a) and (c), and $100.0$ nm for (b) and (d).
The energy is given in units of $t=0.259$ eV.
The horizontal dashed lines denote the Fermi energy $E_F=0.398t=0.103$ eV.
}
\label{energy}
\end{figure}

The red dotted lines in Fig.~\ref{energy} denote the exact energy-dispersion relation
for various situations.
In comparison, the perturbatively obtained energy-dispersion relation
(up to the second order in $\alpha$) ,
\begin{equation}
\label{eq:eigenenergy}
E_{n,i}={\hbar^2k_x^2\over 2m^*}+\mathcal{E}_n+(-1)^{i}\alpha k_x
-{m^*\alpha^2\over2\hbar^2}\, ,
\end{equation}
is also depicted (blue solid lines) in Fig.~\ref{energy}.
Here $\mathcal{E}_n\equiv \hbar^2 k_{n,y}^2/2m^*$
is the energy of the $n$-th ($n$=0,1,2,$\cdots$) excitation mode in the transverse direction,
$k_{n,y}\equiv (n+1)\pi/w$, and $i=1$ or $2$.
Equation~(\ref{eq:eigenenergy}) is obtained by treating the RSO coupling term in $H_{\rm 2D}$
as a perturbation.
Further details on the calculation of the energy dispersion relation via
the second-order degenerate perturbation theory
are given in Appendix~\ref{PerturbationTheory}.

\begin{figure}[b!]
\centerline{\includegraphics[width=4.5cm]{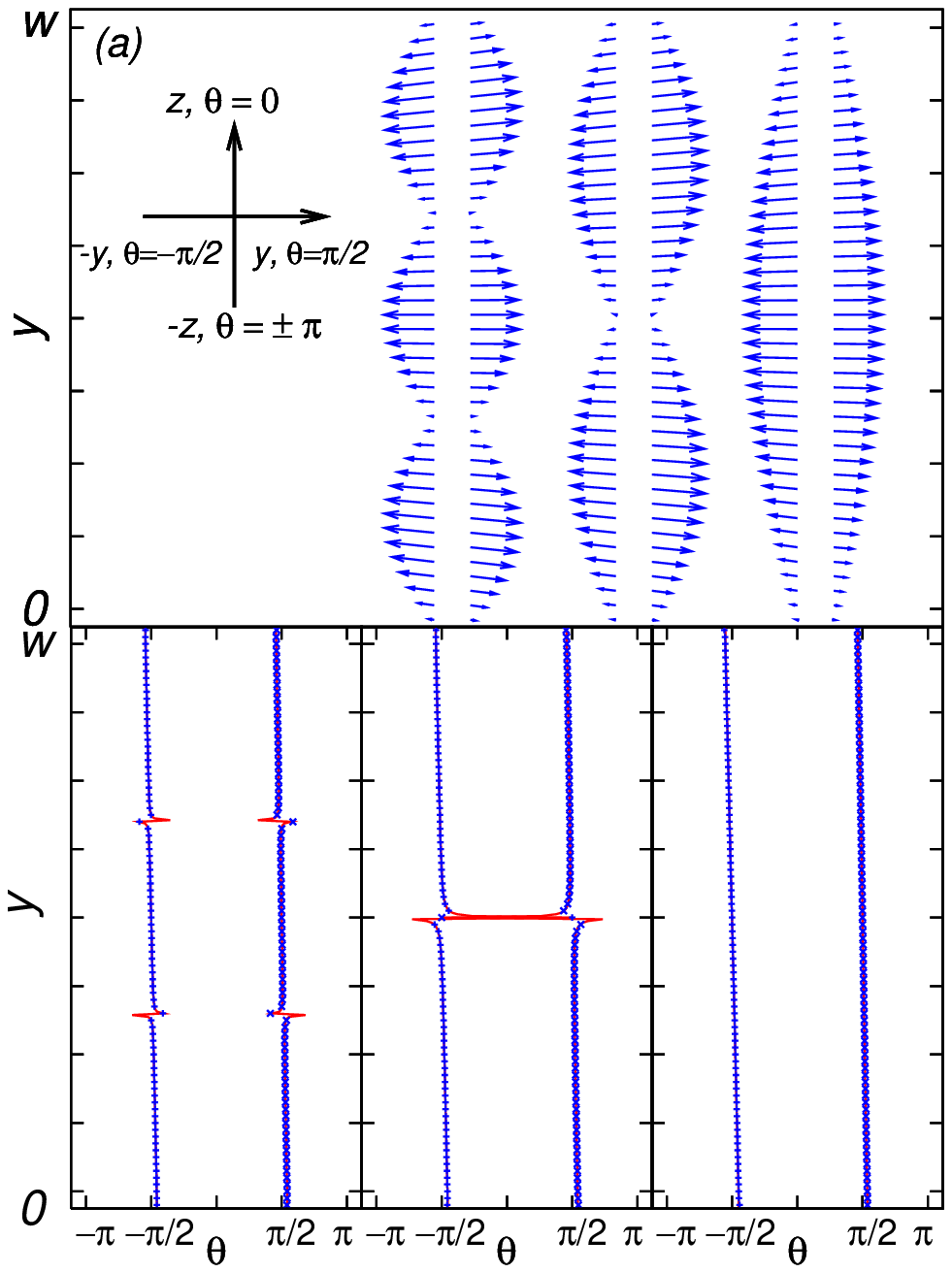}
\includegraphics[width=4.35cm]{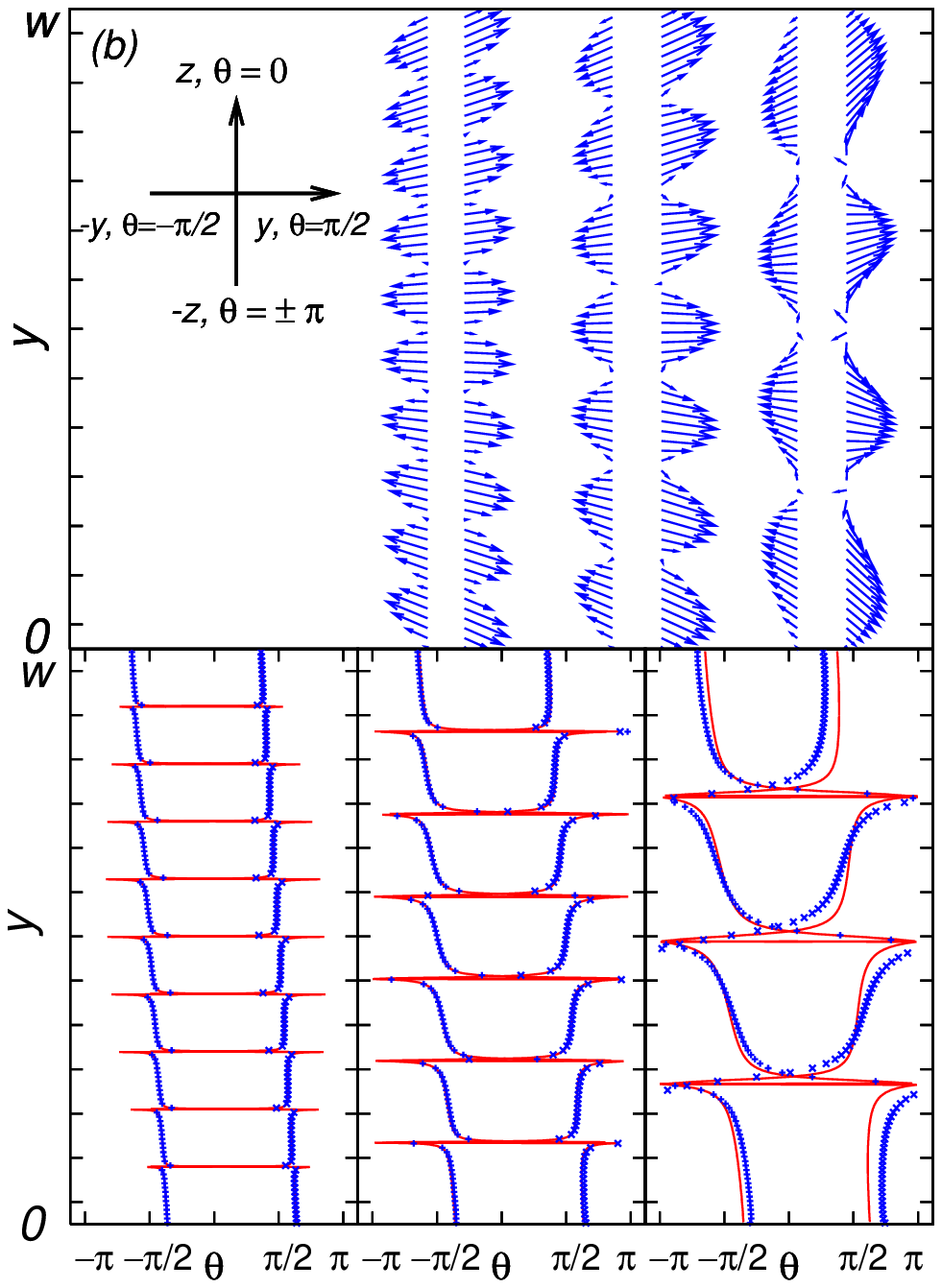}}
\centerline{\includegraphics[width=4.5cm]{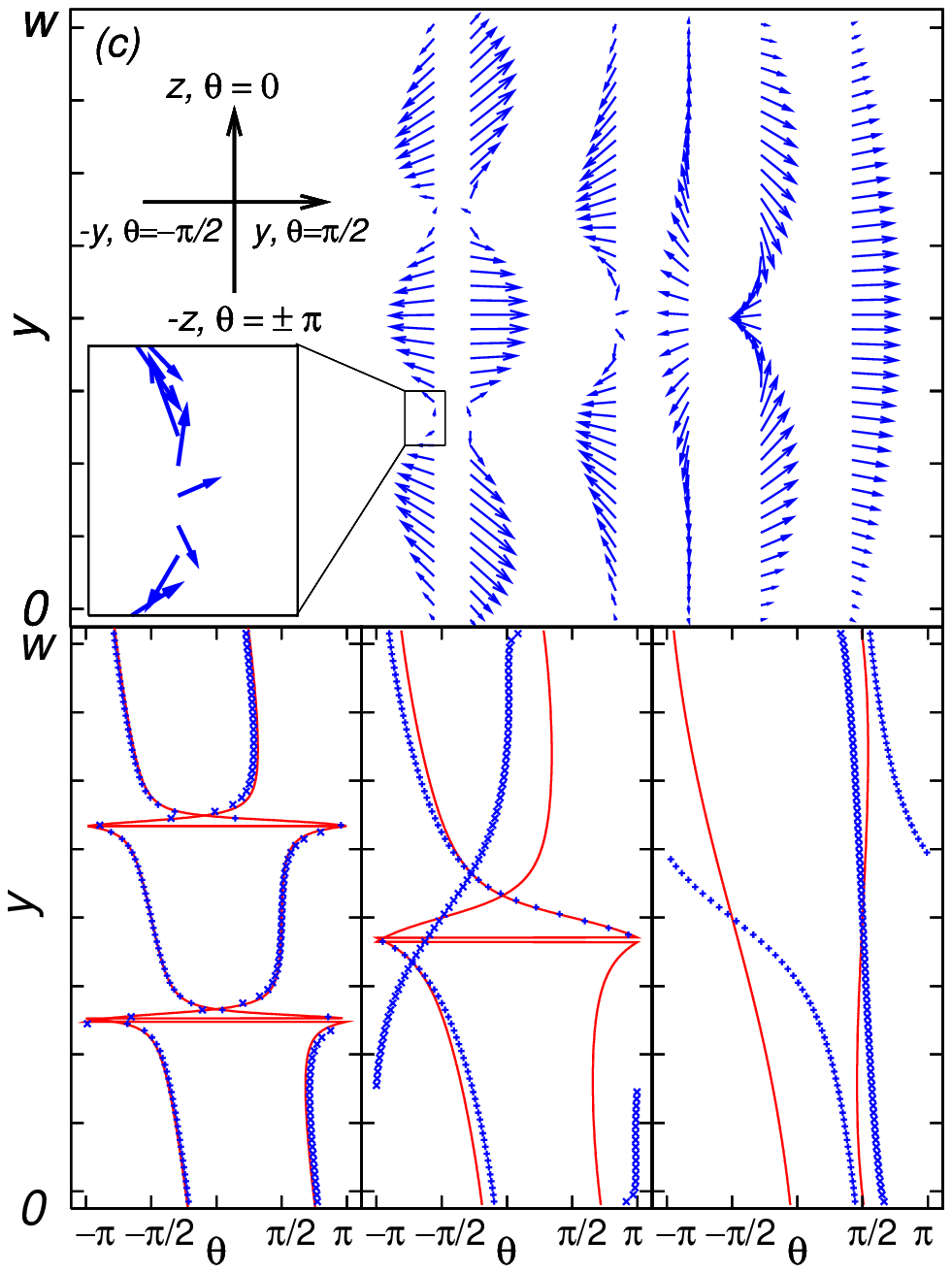}
\includegraphics[width=4.35cm]{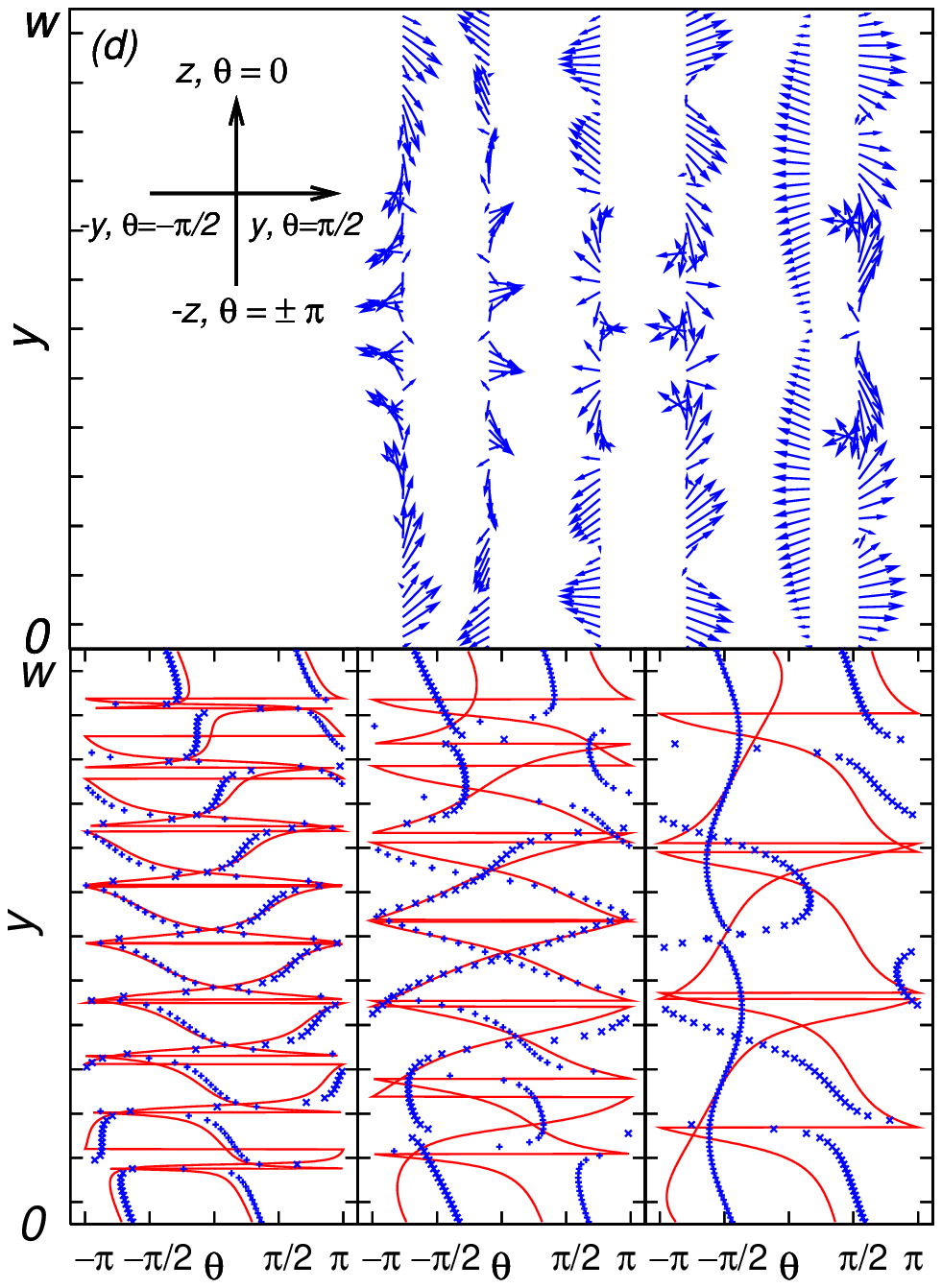}} \caption{(Color
online) Spin texture along the transverse direction for selected
states at $E_F$. The parameters $m^*$, $\alpha$, $w$ and $N_{\rm
ch}$ in (a), (b), (c) and (d) are the same as those in
Figs.~\ref{energy}(a), (b), (c) and (d), respectively.
In the lower panels, the blue dotted lines represent the exact results
while the red solid lines the perturbative results.
Inset in the upper panel of (c) shows the magnified structure
near the points where $\sqrt{\Psi^\dagger(x,y)\Psi(x,y)}$ almost vanishes.
}
\label{SpinTexture}
\end{figure}

Figure~\ref{SpinTexture} shows the local spin directions of the selected eigen wavefunctions
at the Fermi level $E_F$. The width $w$ and the RSO parameter $\alpha$
for Figs.~\ref{SpinTexture}(a), (b), (c), and (d)
are the same as those for Figs.~\ref{energy}(a), (b), (c), and (d), respectively.
The local spin direction of an energy eigen wavefunction is independent of $x$
due to the translational symmetry of $H_{\rm 2D}$ along the $x$-direction.
Also it can be shown that it is always perpendicular to the $x$-axis
and remains within the $yz$-plane
since $H_{\rm 2D}$ remains invariant under the symmetry operator
$\Theta \Pi_x {\cal D}_{\rm spin}(\hat{\bf x},\pi)$
(see Appendix~\ref{Symmetry} for further details),
where $\Theta$ is the time-reversal operator,
$\Pi_x$ is the mirror reflection operator with respect to the $yz$-plane in the orbital space,
and ${\cal D}_{\rm spin}(\hat{\bf x},\pi)$ is the rotation operator with respect to the $x$-axis by the angle $\pi$
in the spin space.
In each upper panel in Figs.~\ref{SpinTexture}(a), (b), (c), and (d),
the arrows indicate the local spin direction as a function of $y$ for six selected exact eigen wavefunctions.
The length of the arrows is proportional to  $\sqrt{\Psi^\dagger(x,y)\Psi(x,y)}$,
which is independent of $x$.
The six states in Figs.~\ref{SpinTexture}(a) and (c) are the six
eigenstates in Figs.~\ref{energy}(a) and (c) at $E_F=0.398
t=0.103$ eV with increasing order of $k_x>0$ . On the other hand,
the six states in Figs.~\ref{SpinTexture}(b) and (d) are the
eigenstates in Figs.~\ref{energy}(b) and (d) at $E_F$ with the
smallest, 2nd smallest, 7th smallest, 8th smallest, 13th smallest,
and 14th smallest $k_x$'s (ordered according to the exact $k_x$
values) . The lower panels in Figs.~\ref{SpinTexture}(a), (b),
(c), and (d) show the local spin angle $\theta$ as a function of
$y$, where $\theta$ is $-\pi/2$, 0, $\pi/2$, $\pm \pi$ when the
local spin direction is pointing $-y$, $+z$, $+y$, $-z$ axis,
respectively. Here $\theta$ is related to the local spin direction
via the spinor representation $(\cos {\theta \over
2}\chi_{\uparrow}+i\sin {\theta \over 2}\chi_{\downarrow})$ of a
local spin direction that lies within the $yz$-plane, and
$\chi_\uparrow=\left( {\tiny \begin{array}{c} 1 \\ 0 \end{array} }
\right)$, $\chi_\downarrow=\left( {\tiny \begin{array}{c} 0 \\ 1
\end{array} } \right)$ are the spinors pointing along the $(\pm
z)$-directions. In each lower panel of Figs.~\ref{SpinTexture}(a),
(b), (c), and (d), there are three subpanels, each of which shows
the profiles of $\theta(y)$'s for two selected energy eigen
wavefunctions; $\theta$'s in the left/middle/right subpanel
correspond to the two leftmost/central/rightmost spin textures in
the upper panel. In the lower panels, blue dotted lines represent
the exact results (see Appendix~\ref{ExactDispersion} for further
details) and red solid lines are from the perturbatively obtained
eigen wavefunctions calculated up to the second order in $\alpha$
(see Appendix~\ref{PerturbationTheory} for further details).

In Figs.~\ref{energy} and \ref{SpinTexture},
two representative values of $\alpha$,
$8.3\times 10^{-12}$ eVm [(a) and (b)] and $49.8 \times 10^{-12}$ eVm [(c) and (d)],
are used.
Figures~\ref{energy}(a) and \ref{SpinTexture}(a) show the result
for the smaller $\alpha$ value and $N_{\rm ch}=3$.
Note that the perturbative result for the energy dispersion
relation is in excellent agreement with the exact result.
Equation~(\ref{eq:eigenenergy}) from the perturbation theory
indicates that the energy dispersion relations for all channel $n$
are identical to each other up to the parallel shift by ${\cal
E}_n$ and remain parabolic. Moveover Fig.~\ref{SpinTexture}(a)
indicates that the spin direction is essentially equal to the
$(+y)$ or $(-y)$-direction, same as in the ideal single-channel
SFET. Thus the SFET for the situation in Fig.~\ref{energy}(a) and
\ref{SpinTexture}(a) should show an almost ideal behavior, which
is indeed the case as demonstrated in Figs.~\ref{multiG01}(a) for
$N_{\rm ch}=3$. The channel-to-channel fluctuations of the spin
precession angle should be negligible.

Figures~\ref{energy}(b) and \ref{SpinTexture}(b) show the result
for $\alpha=8.3\times 10^{-12}$ eVm and $N_{\rm ch}=10$.
The exact energy dispersion relations are again excellently fitted
by the perturbative results [Eq.~(\ref{eq:eigenenergy})], which are
purely parabolic.
The local spin directions, on the other hand,
show noticeable deviations from the ideal $(\pm y)$-directions.
The lower subpanels in Fig.~\ref{SpinTexture}(b) show that
near the ``nodal'' points, where the length of the spin arrows
in the upper panel of Fig.~\ref{SpinTexture}(b) almost vanishes,
$\theta(y)$ changes rapidly by $2\pi$,
forming ``phase-slip''-like structures.
The phase-slip-like structure near the nodal points is more clearly visible in the magnified plot
in the inset of Fig.~\ref{SpinTexture}(c).
Superimposed over the phase-slip-like structures,
there is also a gradual drift of $\theta(y)$ with the increase of $y$.
To understand the origin of the deviations,
the eigen wavefunctions are perturbatively evaluated (see Appendix~\ref{PerturbationTheory}).
Up to the first order in $\alpha$, the perturbative calculation for the reduced wavefunction
$\psi(y)$ [$\Psi(x,y)=e^{ik_x x}\psi(y)$]
results in
\begin{equation}
\label{eq:state01-2}
\psi_{n,i}(y) =
\psi_{n,i}^{(0)}(y)
+(-1)^{i+1}{m^*\alpha(y-w/2)\over\hbar^2}
\, \psi_{n,\bar{i}}^{(0)}(y),
\end{equation}
where $n$ and $i$ are defined in the same way as in Eq.~(\ref{eq:eigenenergy}),
and $\bar{i}=2$ for $i=1$ and $\bar{i}=1$ for $i=2$.
Here the zeroth order reduced wavefunctions are
$
\psi^{(0)}_{n,1}(y)\!=\!\phi_n(y)\cdot
(\chi_\uparrow+i\chi_\downarrow)/\sqrt{2}
$
and
$
\psi^{(0)}_{n,2}(y)\!=\!\phi_n(y)\cdot
(\chi_\uparrow-i\chi_\downarrow)/\sqrt{2}
$
respectively, where $\phi_n(y)\equiv \sqrt{2/w}\, \sin k_{n,y}y$.
From these expressions, one obtains simple expressions for $\theta$,
\begin{eqnarray}
\label{theta_pert}
\theta_{n,i=1}(y)& =& {\pi \over 2}-{2m^*\alpha \over \hbar^2}(y-w/2)
+{\cal O}(\alpha^2), \\
\theta_{n,i=2}(y)& =& -{\pi \over 2}-{2m^*\alpha \over \hbar^2}(y-w/2)
+{\cal O}(\alpha^2), \nonumber
\end{eqnarray}
which has been reported previously~\cite{Hausler01PRB}.
Note that while the zeroth order terms in $\alpha$ are $y$-independent constants, $\pm \pi/2$,
corresponding to the ideal $(\pm y)$-directions,
the first order terms in $\alpha$ are proportional to $y-w/2$,
explaining the gradual drift of $\theta$ with $y$.
In order to gain an insight into the phase-slip-like structure,
we next examine the second order corrections to $\psi(y)$.
Since the first order correction in Eq.~(\ref{eq:state01-2}) identically vanishes
at the nodal points ($\sin k_{n,y}y=0$) of $\psi^{(0)}_{n,i}$,
the second order corrections are the first nonvanishing terms
near those nodal points.
Moreover since $\psi=0$ is a topologically singular point for the local spin angle $\theta$,
even a small and smooth corrections to $\psi$ can result in a rapid change of $\theta$ by $2\pi$.
The local spin angles evaluated from the second order perturbation theory
are shown (red solid lines) in the lower subpanels in Fig.~\ref{SpinTexture}(b),
indeed reproducing the phase-slip-like structure.
The perturbative results are in reasonable agreement with the exact results.
The expressions for the second order corrections to $\psi$ are rather lengthy [Eqs.~(\ref{P2}) and (\ref{Q2})]
and given in the Appendix~\ref{PerturbationTheory} only.
We remark that the gradual drift and the phase-slip-like structures are also
present in Fig.~\ref{SpinTexture}(a), though they are much weaker.
Since they can be explained within the perturbation theory,
the strength of their effects is expected to depend on the magnitude of
the small expansion parameter of the perturbation theory,
which is $2m^*\alpha w/\hbar^2$ according to Eqs.~(\ref{eq:state01-2}), (\ref{P2}), and (\ref{Q2}).
Then recalling that the product $2m^*\alpha w/\hbar^2$ is about 0.28 for
the 3 channel system and about 0.86 for the 10 channel system,
the difference between Figs.~\ref{SpinTexture}(a) and (b) can be explained.
This analysis also implies that the deviations of the local spin configuration
from the ideal spin directions will be magnified with the increase of $N_{\rm ch}$ $(\propto w)$.
Since these deviation will certainly cause the conductance
to deviate from the ideal SFET behavior based on the ideal spin directions,
this analysis also provides an explanation for the initial decay of the conductance modulation ratio
with the increase of $N_{\rm ch}$   demonstrated in Figs.~\ref{multiG01}(a) and \ref{multiG02}(a).

Figures~\ref{energy}(c) and \ref{SpinTexture}(c) show the result
for the larger $\alpha=49.8 \times 10^{-12}$ eVm and $N_{\rm ch}= 3$
with the product $2m^*\alpha w/\hbar^2 \approx$ 1.7,
which is about two times larger than in Figs.~\ref{energy}(b) and \ref{SpinTexture}(b).
Figure~\ref{energy}(c) shows that now some energy dispersion relations
noticeably deviate from the parabolic behaviors.
While many subbands are still well fitted by the parabolic perturbative results,
the two subbands, which are predicted by the perturbation calculation
to cross near $k_x\approx 2.5 \times 10^{8}$ m$^{-1}$,
are affected by the subband mixing and the avoided crossing structure appears,
which goes beyond the perturbation calculation in Appendix~\ref{PerturbationTheory}.
At $E_F$, the states with the second and third largest $k_x$'s are affected by
the avoided crossing and their dispersions deviate from the parabolic dependence,
modifying the spin precession rates of the involved channels as demonstrated recently~\cite{Egues03APL}.
The avoided crossing also affects the local spin angle.
For those two states affected by the avoided crossing,
the exact local spin angle profiles deviate considerably
from the perturbative results [lower subpanels of Fig.~\ref{SpinTexture}(c)].
For the other four states, the exact local spin profiles are
reasonably well fitted by the perturbative results.
Note that the gradual drift of $\theta$ and the phase-slip-like structure
are more evident than in Fig.~\ref{SpinTexture}(b),
which is natural in view of the about two-fold increase of the product $2m^*\alpha w/\hbar^2$.
Based on this analysis, we attribute the reduced conductance modulation ratio
in Figs.~\ref{multiG01}(b) and \ref{multiG02}(b) (even for small $N_{\rm ch}$ such as 3)
jointly to the perturbative spin configuration change and the avoided crossing.
We remark that the avoided crossing in fact occurs in Fig.~\ref{energy}(b) as well,
though not clearly visible in the figure since its effects on the energy dispersions are
restricted to rather narrow ranges of $k_x$.
Furthermore the avoided crossing appears far below
$E_F$, so that its effects on the conductance is negligible.

Figures~\ref{energy}(d) and \ref{SpinTexture}(d) show the result
for $\alpha=49.8 \times 10^{-12}$ eVm and $N_{\rm ch}=10$
with $2m^*\alpha w/\hbar^2 \approx$ 5.2.
It turns out that most subbands are affected by the avoided crossing
and the agreement between the perturbative and exact dispersion relations is poor.
The agreement between the perturbative and exact spin configuration is again poor
for many states at $E_F$.
The abundance of the avoided crossing
explains the strongly suppressed conductance modulation ratio
of the SFET operated with larger $\alpha$ [Figs.~\ref{multiG02}(b) and (c)].

\section{Coherent SFET}
\label{CoherentSFET}
The electronic coherence may give rise to interesting effects in mesoscopic systems~\cite{Datta95Book,Seba01PRL}.
For a single channel system, it has been
predicted~\cite{Schapers01PRB,Mireles02EL,Larsen02PRB,Lee05PRB} that
due to the electron coherence,
the conductance behavior of a SFET deviates from
the conventional sinusoidal dependence $\cos^2(m^*\alpha L/\hbar^2)$.
In this section, we investigate effects of the coherence on multichannel SFETs.
In particular, we focus on situations with $N_{\rm ch}\lesssim 10$ and $\alpha\lesssim 10\times 10^{-12}$ eVm
since for larger $N_{\rm ch}$ and $\alpha$,
the coherence effects will be mixed with
other effects such as nonideal spin configuration and avoided crossing,
making clear identification of the coherence effects difficult.

\begin{figure}[t!]
\centerline{\includegraphics[width=8cm]{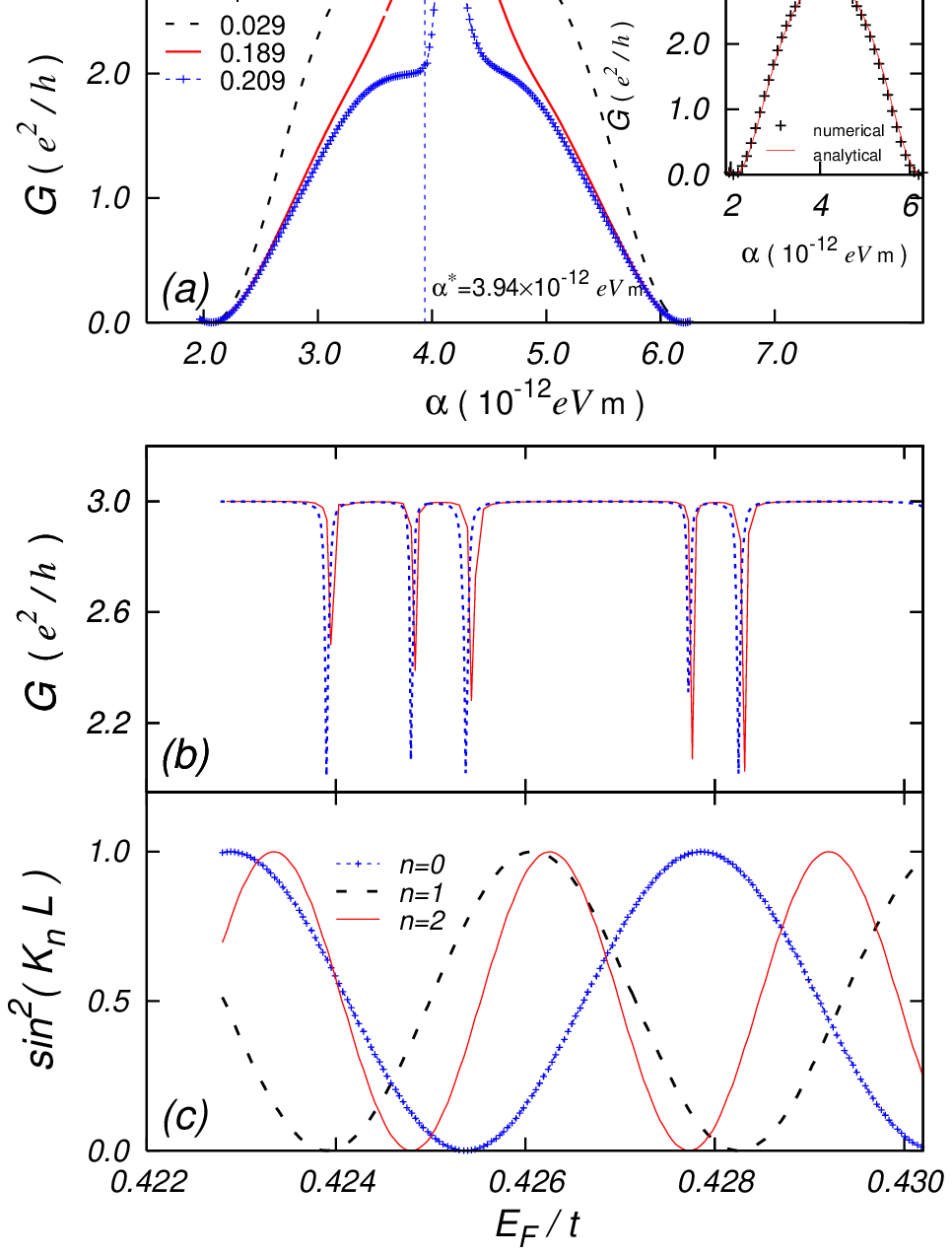}}
\caption{(Color online)
(a)
Variations of the conductance profile
with $E_F$ for a SFET with $N_{\rm ch}=3$ and $L=1.44\,\mu$m at the temperature $T=0$.
$\delta E_F$'s are measured with respect to
the reference energy $0.398t=0.103$ eV.
Inset : The finite temperature conductance $G(\alpha)$ at
$k_BT=10\hbar v_F/L$ and $\delta E_F/t=0.209$
where $v_F\equiv \hbar(K_0+K_1+K_2)/3m^*$.
Note that the numerical result is in excellent agreement with
the analytic high temperature expression~\cite{ThermalAverage}.
(b) The red solid line represents the numerically calculated
$G(\alpha^*=3.94\times 10^{-12}\, {\rm eVm})$ as a function of $E_F$ at $T=0$.
Here $\alpha^*$ [vertical dotted line in (a)] is close to the conductance maximum point.
Thus dips in the $G(\alpha^*)$ graph imply the appearance of the nested peak structure
at the corresponding $E_F$.
For comparison, the prediction of the formula [Eq.~(\ref{eq:G1})] is
also shown (blue dashed line).
(c) $\sin^2(K_nL)$ for $n=$0, 1, and 2.
The comparison between (b) and (c) indicates that
the dip positions of $G(\alpha^*)$ coincide with the points
where one of $\sin^2(K_nL)$'s becomes zero.
}
\label{peak05}
\end{figure}

Hints of the coherence effects appear already at Fig.~\ref{multiG01}(a).
A close look reveals the deviation of the line profile $G(\alpha)$
from the sinusoidal behavior $\propto \cos^2(m^*\alpha L/\hbar^2)$.
The evolution of $G(\alpha)$ with the variation of the Fermi energy $E_F$
illustrates the deviation more clearly.
Figure~\ref{peak05}(a) shows $G(\alpha)$ of a SFET with $N_{\rm ch}=3$
for a number of $E_F$'s.
$\delta E_F$'s in Fig.~\ref{peak05}(a) are measured with respect to
the reference energy
0.103 eV$=$0.398$t$.
While $G(\alpha)$ is not sensitive to $E_F$
near the conductance maximum (near $\alpha=4.2\times 10^{-12}$ eVm)
and the minimum (near $\alpha=2.2\times 10^{-12}$ eVm and $6.2\times 10^{-12}$ eVm),
the profiles of $G(\alpha)$ including the full-width-at-half-maximum (FWHM)
exhibit considerable dependence on $E_F$.
Moreover near certain special $E_F$'s such as
$\delta E_F/t=0.209$,
$G(\alpha)$ shows a nested peak structure,
where a small peak appears on top of a bigger background peak.
These results are in clear contrast to the result in Ref.~\cite{Mireles01PRB},
which reports the absence of the $E_F$-dependence when $2m^*\alpha w/\hbar^2 \ll 2\pi$.
While our conductance calculation takes into account all three parts
(2DEG, injector, and collector) quantum mechanically,
only the 2DEG is quantum mechanically treated in Ref.~\cite{Mireles01PRB}.
We believe that this is responsible for the difference.

In order to examine the connection between the electronic coherence and the nested peak structure,
we first recall that in a single channel SFET,
the conductance calculation~\cite{Lee05PRB}, which takes into account the electronic coherence,
results in
\begin{equation}
G={e^2\over h}
{4\cos^2{\phi_{\alpha}\over 2}
\sin^2(K L)\over
\sin^4{\phi_\alpha\over 2}
+4\cos^2{\phi_\alpha\over 2}
\sin^2(KL)}\, ,
\label{eq:single-channel}
\end{equation}
where $\phi_{\alpha}\equiv 2m^{*}\alpha L/\hbar^2$,
$K=[(k_x)_{i=1}+(k_x)_{i=2}]/2$, and
$(k_x)_{i=1}$ and $(k_x)_{i=2}$ $(>0)$ are the two longitudinal wavevectors  at $E_F$.
The deviation of Eq.~(\ref{eq:single-channel})
from the sinusoidal behavior arises from the fact that
the injector and collector behave as reflecting walls
for the electron spin component antiparallel to
the favored spin direction in the injector and collector.
Thus the unfavored spin component may be reflected many times by the ideal injector and collector
until it acquires, via the spin precession by the RSO coupling,
some component parallel to the favored spin component
and is transmitted to the injector or collector.
Thus in addition to the direct contribution that does not undergo any reflection,
there are
$n$-th order contributions to the electron transmission that undergo
the reflection $2n$ times until being transmitted to the collector.
In this sense, the coherent SFET is analogous (though not identical) to the Fabry-Perot interferometer.
It is straightforward to verify that
the amplitude of the $(n+1)$-th contribution is smaller than
that of the $n$-th contribution
by the factor $-\exp(i2KL)\sin^2(\phi_\alpha/2)$ for $n=0$
and by the factor $\exp(i2KL)\cos^2(\phi_\alpha/2)$ for $n\ge 1$.
Here the zeroth ($n=0$) contribution denotes the direct contribution.
Note that the interference between
the direct contribution and the multiply reflected contributions is destructive (constructive)
when the phase $\exp(i2KL)$ is $+1$ ($-1$).
This explains the factor $\sin^2(KL)$ in Eq.~(\ref{eq:single-channel}),
which predicts the suppression (enhancement) of the FWHM of the profile $G(\alpha)$
when $\sin^2(KL)\approx 0$ (1).
Since $K$ depends on $E_F$, this $K$ dependence also implies the $E_F$ dependence.
To examine the coherence effects in a multichannel SFET,
we use an intuitive generalization of Eq.~(\ref{eq:single-channel}) to a multichannel system;
\begin{equation} \label{eq:G1}
G
={e^2\over h}\sum_{n=0}^{N_{\rm ch}-1}g_n
={e^2\over h}\sum_{n=0}^{N_{\rm ch}-1}
{4\cos^2{\phi_{\alpha}\over 2}
\sin^2(K_nL)\over
\sin^4{\phi_\alpha\over 2}
+4\cos^2{\phi_\alpha\over 2}
\sin^2(K_nL)}\, ,
\end{equation}
where $K_n\equiv[(k_{x})_{n,i=1}+(k_{x})_{n,i=2}]/2$,
and $(k_{x})_{n,i}$ $(>0)$ is the longitudinal wavevector at $E_F$
for the subband $(n,i)$ (see Fig.~\ref{energy}).
Equation~(\ref{eq:G1}) is expected to be a reasonable approximation
when the inter-channel mixing is weak and each channel is close to an ideal 1D channel,
which are indeed the case
for $\alpha\lesssim 10\times 10^{-12}$ eVm and $N_{\rm ch}\lesssim 10$
as demonstrated in Sec.~\ref{conductance}.
In Fig.~\ref{peak05}(b),
the prediction of Eq.~(\ref{eq:G1}) (blue dashed line)
for $G(\alpha^*)$ as a function of $E_F$ is compared with
the numerical TB conductance result (red solid line),
where $\alpha^*=3.94\times 10^{-12}$ eVm [vertical dashed line in Fig.~\ref{peak05}(a)].
Since $\alpha^*$ is close to the position where the conductance retains its maximal value $(e^2/h)N_{\rm ch}$,
the dips in $G(\alpha^*)$ imply the formation of the nested peak structure.
To compare with the TB calculation result,
$K_n$ and $L$ in Eq.~(\ref{eq:G1}) are expressed in terms of
the TB parameters.
To evaluate $K_n=[(k_x)_{n,i=1}+(k_x)_{n,i=2}]/2$, we use the relation
$E_F=2t\left\{2-\cos[(k_x)_{n,i}a]-\cos[(n+1)\pi a/ w]\right\}
+(-1)^i(\alpha/a)\sin[(k_x)_{n,i}a]-m^*\alpha^2/2\hbar^2$,
which amounts to the TB approximation of Eq.~(\ref{eq:eigenenergy}).
Here $w=a(N_y+1)$.
To evaluate $L$, we use $L=a(N_x+1+\beta)$,
where $\beta=0.6$ is an {\it ad hoc} fitting parameter
much smaller than $N_x=749$.
Note that Eq.~(\ref{eq:G1}) produces dips (blue dashed line)
and moreover the dip positions predicted by Eq.~(\ref{eq:G1}) are
in good agreement with the TB conductance calculation result (red solid line).
Thus Eq.~(\ref{eq:G1}) reproduces the nested peak structure.
We remark that when $\beta=0$ is used, the dips in the blue dashed line shift to the right
by $0.0001\sim 0.0003t$ only (not shown),
not affecting the agreement significantly.

The agreement provides a simple insight into the origin of the nested peak structure.
A key observation is that while all $g_n$'s in Eq.~(\ref{eq:G1}) produce the conductance peaks
at the same $\alpha$, the peak widths may fluctuate considerably from channel to channel.
The channel-to-channel peak width fluctuation arises from the dependence of the peak width
on $K_n$.
In particular when $\sin^2(K_n L)\approx 0$, the peak produced by $g_n$ becomes very narrow.
Thus when $\sin^2(K_n L)$ is close to zero for one particular $n$,
one very narrow peak for that particular $n$ is superposed with the other broad peaks,
generating the nested peak structure.
Figure~\ref{peak05}(c) plots $\sin^2(K_n L)$'s as a function of $E_F$.
Note that whenever one of $\sin^2(K_n L)$ vanishes,
a dip appears in Fig.~\ref{peak05}(b).
We remark that though the nested peak structure is illustrated mainly for $N_{\rm ch}=3$,
it appears for higher $N_{\rm ch}$ as well.
An example for $N_{\rm ch}=20$ is visible in Fig.~\ref{multiG01}(a)
near $\alpha=4.3\times 10^{-12}$ eVm.
For sufficiently large $N_{\rm ch}$, however,
the nested peak structure is expected to vanish.
The nested peak structure was not found in a two-dimensional coherent SFET~\cite{Matsuyama02PRB},
which amounts to the $N_{\rm ch}\rightarrow \infty$ limit.

\begin{figure}[b!]
\centerline{\includegraphics[width=8cm]{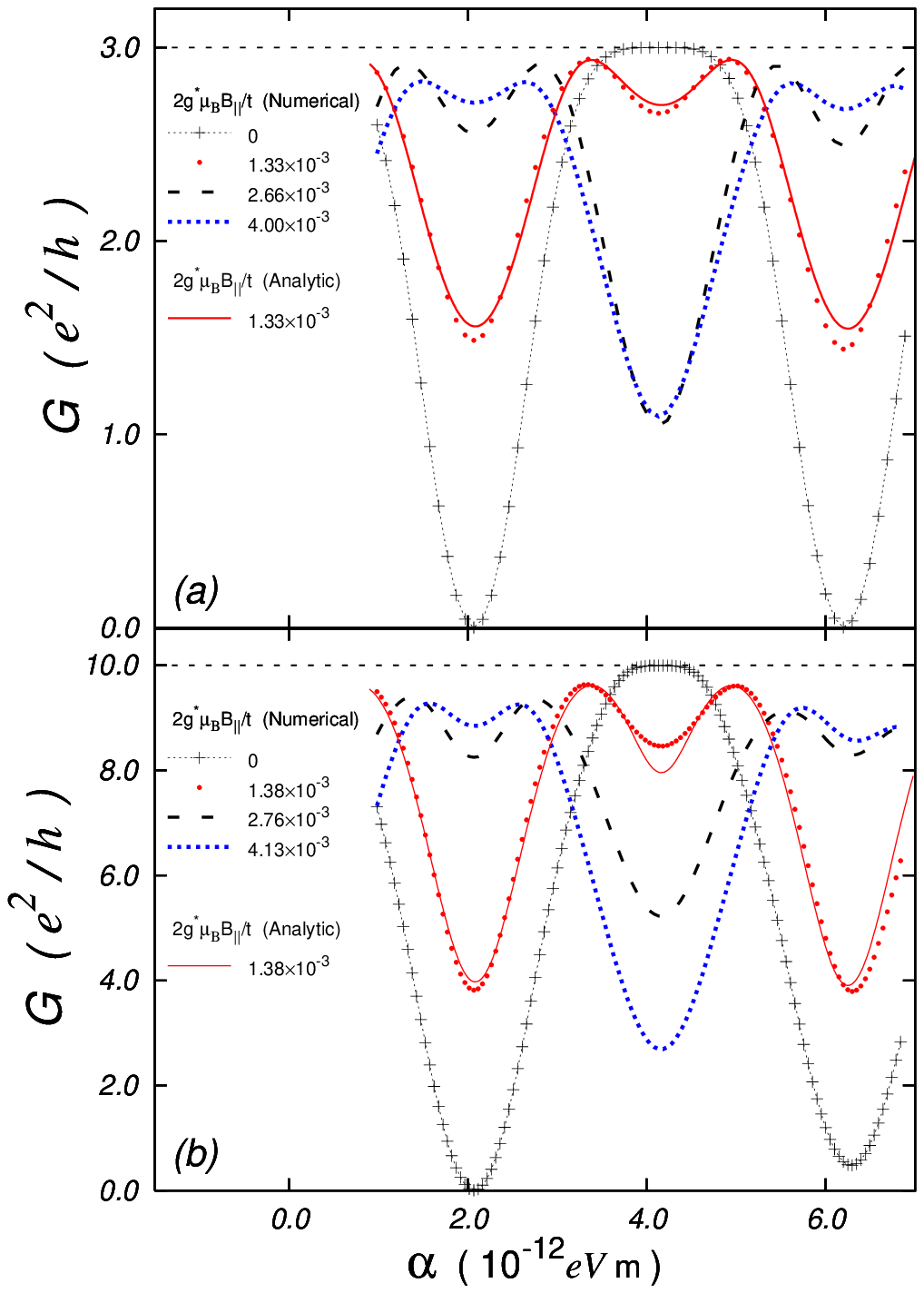}}
\caption{(Color online) Conductance $G$
as a function of the RSO coupling parameter $\alpha$
for (a) $N_{\rm ch}=3$ and (b) $N_{\rm ch}=10$.
The solid lines in (a) and (b) show the analytic results obtained using Eq.~(\ref{eq:G2})
for a specific field strength $2g^*\mu_BB_{\parallel}/t=1.33\times 10^{-3}$
and $1.38\times 10^{-3}$, respectively.
Note that the analytic results are in good agreement with the numerical results.
}
\label{mag}
\end{figure}

The electronic coherence can give rise to
another interesting effect
when an external magnetic field $\vec{B}$ is applied,
providing an additional source of the spin precession.
Here we consider $\vec{B}=B_\parallel \hat{y}$ applied along the $y$-axis,
so that the spin precession axis by $\vec{B}$ is parallel
to that due to the RSO coupling in the ideal 1D situation.
The spin precessions by $B_\parallel$ and the RSO coupling differ in one crucial respect;
When the electron reverses its motion,
the direction of the spin precession by the RSO coupling is reversed
while that of the spin precession by $B_\parallel$ is not.
For a single channel SFET, it is illustrated~\cite{Lee05PRB} that
this difference, combined with the coherent Fabry-Perot-like interference,
gives rise to the splitting of the conductance peaks upon the application of $B_\parallel$.
Figure~\ref{mag} shows that the magnetic-field-induced peak splitting
occurs also for multichannel SFETs.
The numerical TB conductance calculation results are compared with the expression,
\begin{widetext}
\begin{equation} \label{eq:G2}
G={e^2\over h}\sum_{n=0}^{N_{\rm ch}-1}g_n
={e^2 \over h}\sum_{n=0}^{N_{\rm ch}-1}
{4(\cos^2{\phi_\alpha\over 2}
-\sin^2{\phi_{B,n}\over 2})\sin^2K_n L
+4\sin^2{\phi_\alpha\over 2}\sin^2{\phi_{B,n}\over 2}\over
(\sin^2{\phi_\alpha\over 2}\!
+\sin^2{\phi_{B,n}\over 2})^2
+4(\cos^2{\phi_\alpha\over 2}\!
-\sin^2{\phi_{B,n}\over 2})\sin^2 K_n L},
\end{equation}
\end{widetext}
which is an intuitive extension of the single channel formula in Ref.~\cite{Lee05PRB}.
Here $\phi_{B,n}$ denotes the spin precession angle due to $B_\parallel$.
For weak fields, it is given by $2g^*\mu_B B_\parallel/(\hbar v_{F,n}/L)$,
where $g^*$ is the effective Lande-$g$ factor,
$\mu_B$ is the Bohr magneton,
and $v_{F,n}=\hbar K_n/m^*$ is the group velocity of electrons with energy $E_F$
in the $n$-th channel~\cite{Kn-comment}.
Note that the predictions of Eq.~(\ref{eq:G2}) are
in reasonable agreement with the numerical TB results both for $N_{\rm ch}=3$ and 10.

\section{Discussion}
\label{nonidealSFET}
An ideal spin injector/collector has been assumed in the preceding sections
in order to focus on the multichannel effects.
For a certain material to be an ideal spin injector/collector,
(i) its spin polarization should be 100\%
and
(ii) its effective mass and Fermi wavelength should match those for
the 2DEG. While the condition (i) is satisfied in the
so-called half-metals~\cite{Groot83PRL} and the condition (ii) may be satisfied
in ferromagnetic semiconductors or diluted magnetic semiconductors~\cite{Ohno98Science},
we are not aware of any material that satisfies the both conditions.
In Sec.~\ref{nonidealInjCol}, we thus address briefly effects of
a \textit{nonideal} spin injector/collector on the multichannel SFET.
%
\subsection{Nonideal spin injector/collector}
\label{nonidealInjCol}
%
One of most explored and representative choices for a spin injector/collector is
conventional metallic ferromagnets, whose spin polarization
can be as high as $\sim50\%$.
Use of conventional metallic ferromagnets requires a special care.
Recent theories~\cite{Rashba00PRB} proposed that the spin injecton/detection
rate can be greatly improved by introducing a thin
insulator between a conventional ferromagnet and the 2DEG
while without the insulator, the spin injection/detection rate
was demonstrated~\cite{Filip00PRB} to be below 1\% due to the so-called conductance mismatch
problem~\cite{Schmidt00PRB}.
The physics behind the insulator ``prescription'' for the conductance mismatch
problem is that the tunneling through the insulating barrier becomes
spin-dependent~\cite{Rashba00PRB} in the ferromagnet-insulator-2DEG geometry even though
the insulator is nonmagnetic.
Subsequent experiments~\cite{Motsnyi03PRB} using oxide insulating barriers have reported
the enhanced spin polarization of 2-30\% at room temperatures.
Schottky tunnel barriers were also demonstrated to be effective
and the spin polarization of 2-30\% has been reported~\cite{Zhu01PRL} at low temperatures.
%
\begin{figure}[b!]
\centerline{\includegraphics[width=8cm, height=10cm]{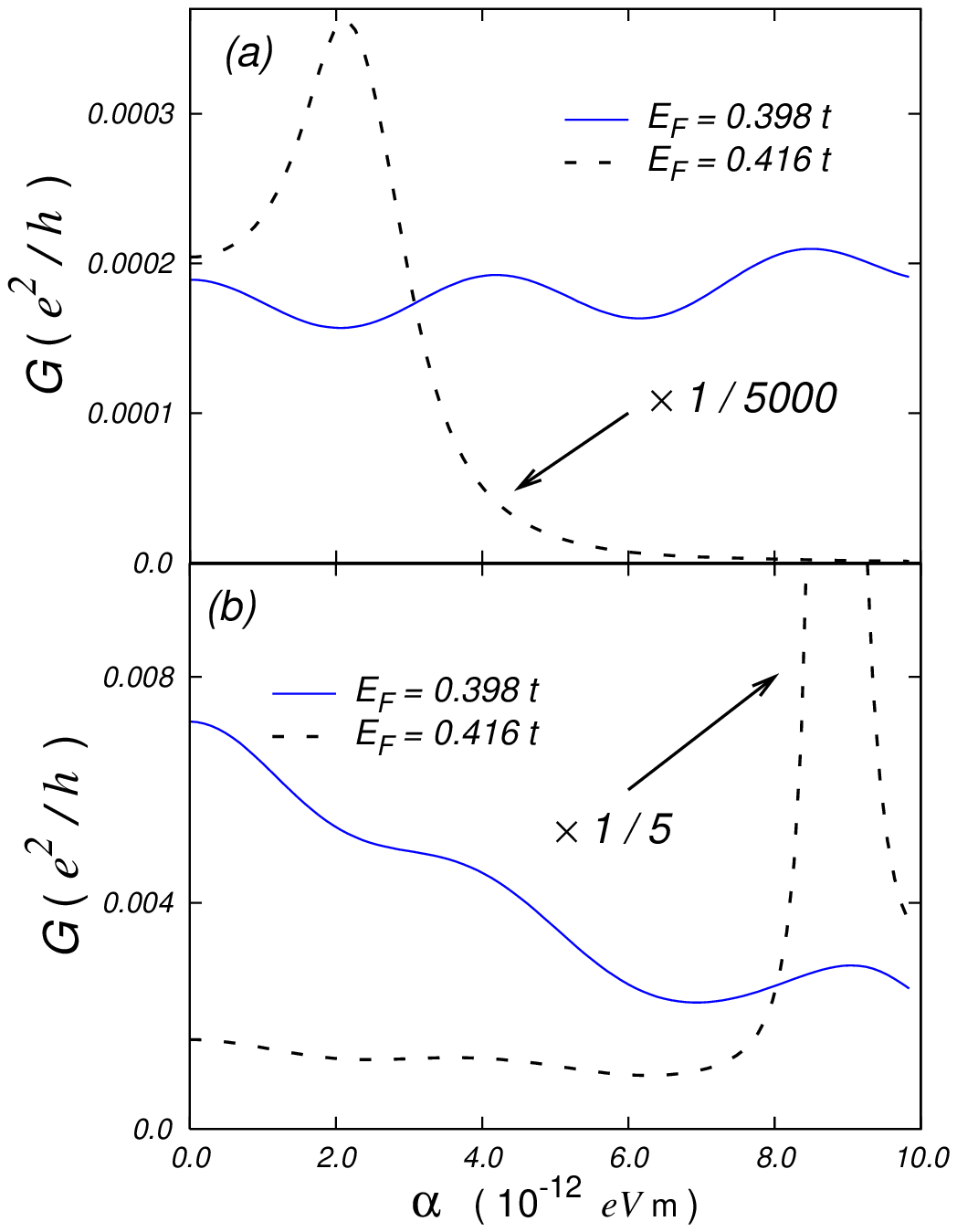}}
\caption{(Color online) The conductance $G$ of a ballistic SFET
with the nonideal injector/collector
described by the modified
$H_{\rm coupling}^{\rm TB}$[Eq.~(\ref{eq:couplingTB2})]
and
$H_{\rm inj/col}^{\rm TB}$.
$N_{\rm ch}=3$ in (a) and
$N_{\rm ch}=10$ in (b).
The length $L$ of the 2DEG is $1.44\,\mu$m in both (a) and (b).
Other parameters are the same as those in Fig.~\ref{multiG01}.
The spin polarization of the injected current
for the ferromagnet-insulator-2DEG structure is assumed to be about 30\%.
}
\label{non1}
\end{figure}
%


Here we calculate numerically the conductance of the multichannel SFET
equipped with the nonideal spin injector/collector that consists of
a conventional metallic ferromagnet-thin insulator-2DEG hybrid
structure.
%
\begin{figure}[t!]
\centerline{\includegraphics[width=8cm, height=9.5cm]{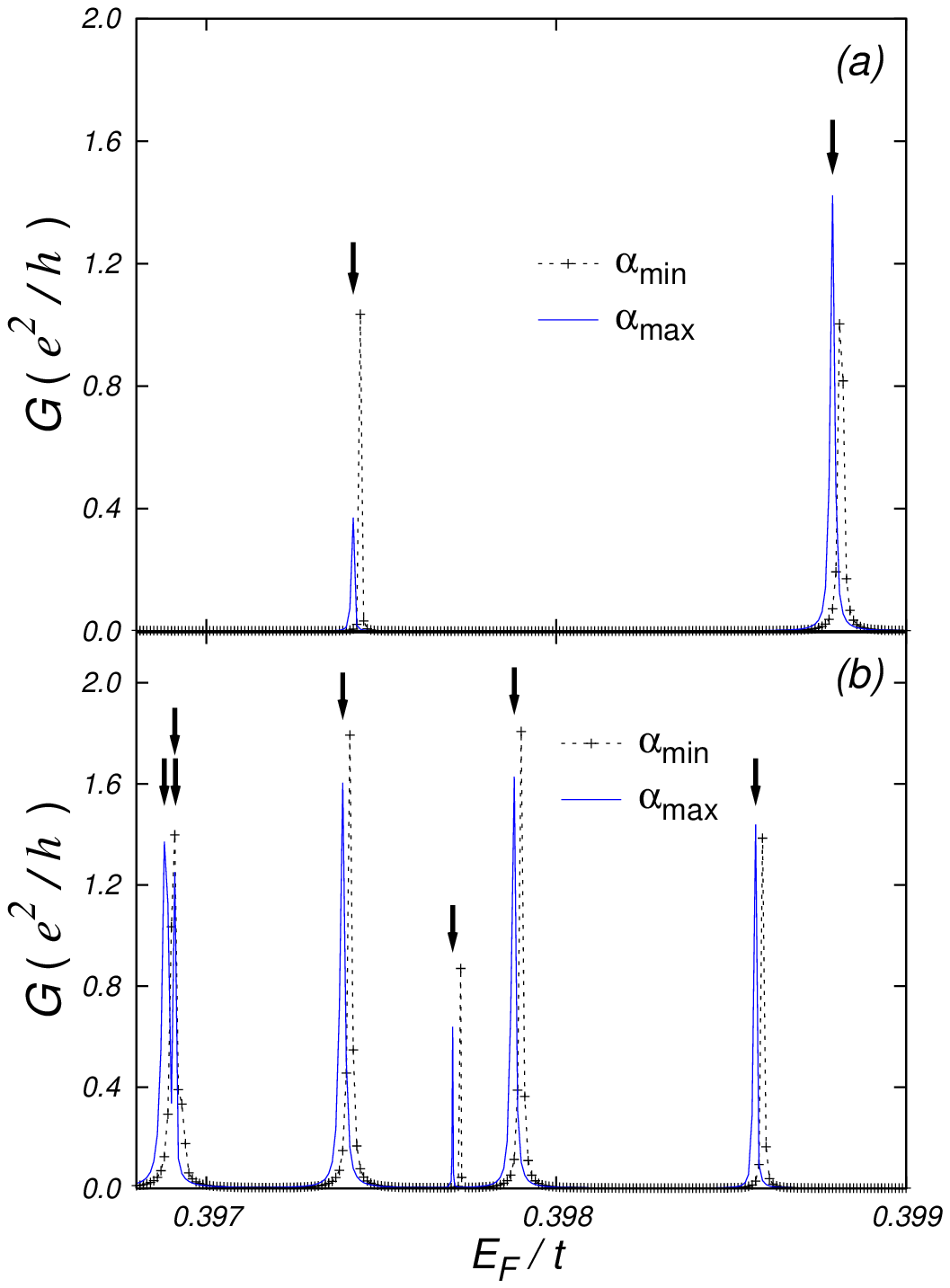}
}
\caption{(Color online) The conductance $G$ as a function of
the Fermi energy $E_F$ with fixed $\alpha$, $\alpha_{\rm min}\equiv 2.07\times 10^{-12}$ eVm
and $\alpha_{\rm max}\equiv 4.13\times 10^{-12}$ eVm, for $N_{\rm ch}=3$ in (a)
and
$N_{\rm ch}=10$ in (b).
The peak positions of $G$  for $\alpha_{\rm max}$ are marked by arrows.
Note that there are three closely spaced peaks near $E_F=0.397t$.
The length $L$ of the 2DEG is $1.44\,\mu$m.
}
\label{non2}
\end{figure}
%
One possible way to model the spin-dependent tunneling in the ferromagnet-insulator-2DEG structure is
to explicitly take into account all three materials, ferromagnet,
insulator, and 2DEG (Appendix~\ref{Injection}).
This approach is however too demanding for numerical calculations
since the Fermi wavelength in the ferromagnet is
about two orders of magnitude smaller than that in the 2DEG.
We thus adopt a simplified phenomenological description,
where all effects of the ferromagnet
(including not only its spin-dependent density of states but also
the effective mass and the Fermi wavelength differences from
those of the 2DEG)
are taken into account by phenomenological spin-dependent hopping parameters
for the tunneling through the insulator.
To be specific, the TB Hamiltonian [Eq.~(\ref{eq:couplingTB})]
for the coupling between the 2DEG and the injector/collector
is modified as follows:
%
\begin{eqnarray}
\label{eq:couplingTB2}
H_{\rm coupling}^{\rm TB}&=
- \sum_{s_x}\sum_{j=1}^{N_y}
t_{s_x}\left( c_{1,j,s_x}^\dagger c_{0,j,s_x}\right.\nonumber \\ 
&\left.+c_{N_x+1,j,s_x}^\dagger c_{N_x,j,s_x}+ {\rm H.c.} \right),
\end{eqnarray}
where $t_{s_x}$ is a spin-dependent hopping parameter that
takes into account not only the spin-dependent tunneling through
the insulator but also other effects from the effective mass and the Fermi
wavelength differences between the ferromagnet and the 2DEG.
Since all effects of the ferromagnet are already
taken into account in $t_{s_x}$ [Eq.~(\ref{eq:couplingTB2})],
the injector/collector may be modelled as a \textit{nonmagnetic} semiconducting lead,
which can be described by modifying
$H_{\rm inj/col}^{\rm TB}$ [Eqs.~(\ref{eq:InjTB}) and~(\ref{eq:ColTB})] so that
similar contributions for $s_x=-1$ are also included.
Although this description adopts a considerably simplified picture of the real system,
it is still expected to capture the main physics of the spin injection/detection
since, as demonstrated in Refs.~\cite{Rashba00PRB} and Appendix~\ref{Injection},
the spin-dependent tunneling through the insulator
is the main source of the sizable
spin injection/detection rate ($\sim30\%$) while the bulk properties of the ferromagnet
generate much weaker contributions only.
We choose the value of $t_{s_x}$ to be $0.1t$ for $s_x=1$ and $0.073t$ for
$s_x=-1$, respectively. The resulting spin
polarization of the injected current
estimated by
$(t^2_{s_x=1}-t^2_{s_x=-1})/(t^2_{s_x=1}+t^2_{s_x=-1})$ is then 30\%,
which is similar to the values reported in experiments~\cite{Motsnyi03PRB}.
%
\begin{figure}[t!]
\centerline{\includegraphics[width=8cm]{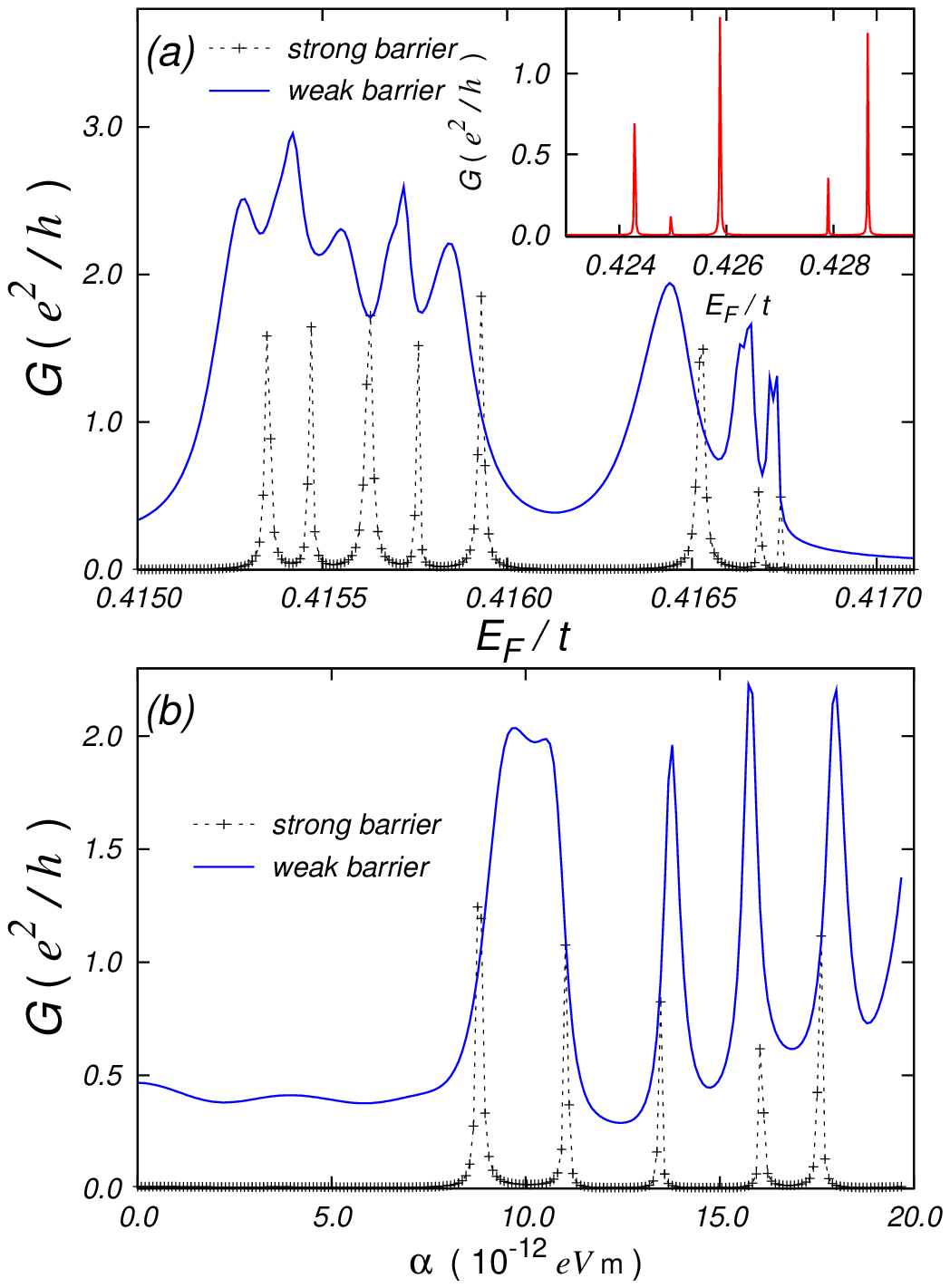}}
\caption{(Color online) (a) The conductance $G$ as a function of $E_F$
for the SFET with two nonideal spin injector/collector that
contains a thin insulating barrier.
Here $N_{\rm ch}=10$ and $\alpha=8.3\times10^{-12}$ eVm.
For the curves labelled as the strong and weak barriers,
the hopping parameters [Eq.~(\ref{eq:couplingTB2})] are
($t_{s_x=1}\!=\!0.1t$, $t_{s_x=-1}\!=\!0.073t$)
and
($t_{s_x=1}\!=\!3\!\times\!0.1t$, $t_{s_x=-1}\!=\!3\!\times\!0.073t$), respectively.
Inset in (a) : $G(E_F)$ for $N_{\rm ch}=3$.
Note that the positions of the large resonance peaks
nearly coincide with those of the nested peaks in Fig.~\ref{peak05}(b).
(b) The conductance $G$ as a function of $\alpha$
for the strong and weak barriers when $N_{\rm ch}=10$ and $E_F=0.416t$.
The length $L$ of the 2DEG is $1.44\,\mu$m.
}
\label{non3}
\end{figure}
%

For $N_{\rm ch}=3$ and $E_F=0.398t$ [solid line in Fig.~\ref{non1}(a)],
$G$ clearly shows the modulation by $\alpha$
though the value of $G$ is considerably reduced below $0.001\times(e^2/h)$
due to the insulating barriers
and the ratio $G_{\rm max}/G_{\rm min}\simeq1.7$ is also
suppressed due to the reduced spin injection/detection rate of 30\%.
Thus for this particular situation,
$G$ shows rather expected behaviors.
For other situations, somewhat unexpected behaviors also appear.
For $N_{\rm ch}=3$ and $E_F=0.416t$ [dashed line in Fig.~\ref{non1}(a)],
the ``periodic'' modulation disappears
and instead a large peak appears at $\alpha\simeq2.1\times10^{-12}$ eVm
with the maximum conductance of $\sim1.8\times(e^2/h)$
(the peak height is scaled down by the factor of 5000 to fit in the vertical
scale of the graph).
For $N_{\rm ch}=10$ and $E_F=0.416t$ [dashed line in Fig.~\ref{non1}(b)],
a large peak (near $\alpha\simeq8.9\times10^{-12}$ eVm) coexists with the
$G$ modulation in the lower $\alpha$ range.
For $N_{\rm ch}=10$ and $E_F=0.398t$ [solid line in Fig.~\ref{non1}(b)],
the $G$ modulation is superimposed on a background slope.

%
\begin{figure}[t!]
\centerline{\includegraphics[width=5cm, angle=-90]{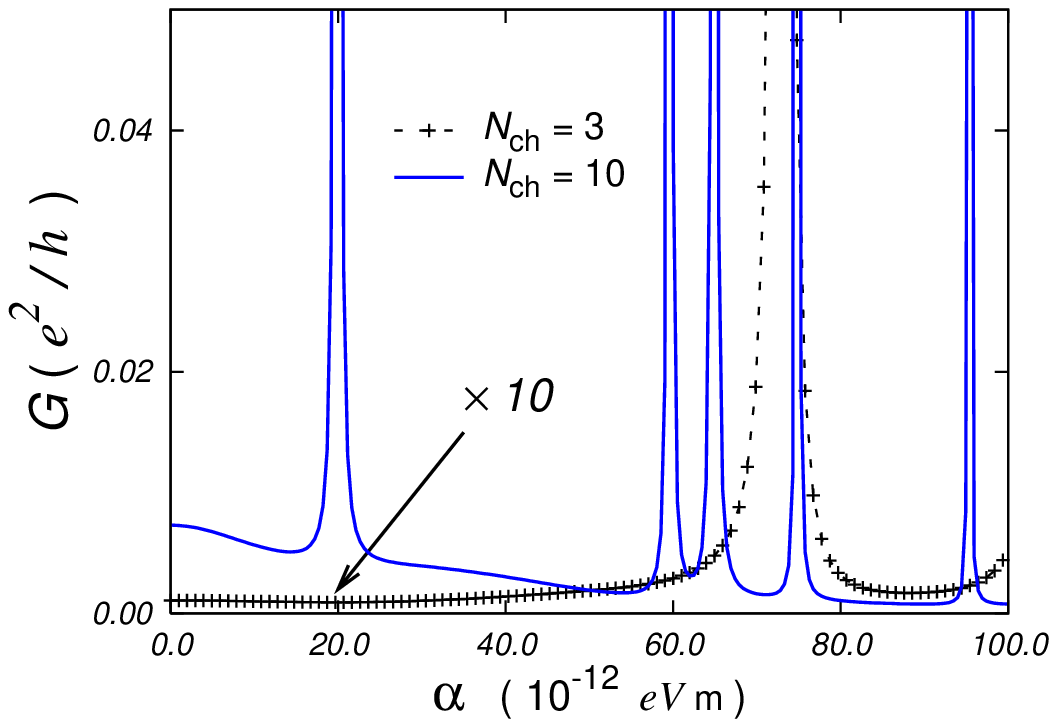}}
\caption{(Color online) The conductance $G$ of a ballistic SFET
with the nonideal injector/collector
as a function of the RSO coupling parameter $\alpha$ at $T=0$.
The length $L$ of the 2DEG is $0.1\times1.44\,\mu$m and $E_F=0.398t=0.103$ eV.
}
\label{non4}
\end{figure}

To investigate how frequently such peaks appear,
$G$ is evaluated as a function of $E_F$ for two representative values of $\alpha$,
$\alpha_{\rm min}\equiv 2.07\times 10^{-12}$ eVm and $\alpha_{\rm max}\equiv 4.13\times 10^{-12}$ eVm,
which correspond to the positions of the local
minimum and maximum of $G$ at $E_F=0.398t$ for $N_{\rm ch}=3$.
For $N_{\rm ch}=3$ [Fig.~\ref{non2}(a)],
two peaks appear in the depicted $E_F$ range
and for $N_{\rm ch}=10$ [Fig.~\ref{non2}(b)], seven peaks appear.
Thus the occurrence of the peaks becomes more frequent with the increase of $N_{\rm ch}$.
The calculation for $N_{\rm ch}$=3, 10, and 20 over a much wider range of $E_F$
($0.360t<E_F<0.440t$, not shown) verifies this trend.
Note that the peak positions for
$\alpha_{\rm min}$ and $\alpha_{\rm max}$ are
somewhat different, implying that the peak positions shift with $\alpha$.
Considering that the peak heights ($\sim e^2/h$) are three or four
orders of magnitudes larger than the modulation amplitudes in Fig.~\ref{non1},
even tails of the peaks can be large ``perturbations'' to the modulation
and thus the $\alpha$-dependence of the peak positions, though weak,
can affect the modulation considerably.
As remarked above, this perturbation appears more frequently for larger $N_{\rm ch}$.
%
\begin{figure}[t!]
\centerline{\includegraphics[width=8cm, height=10cm]{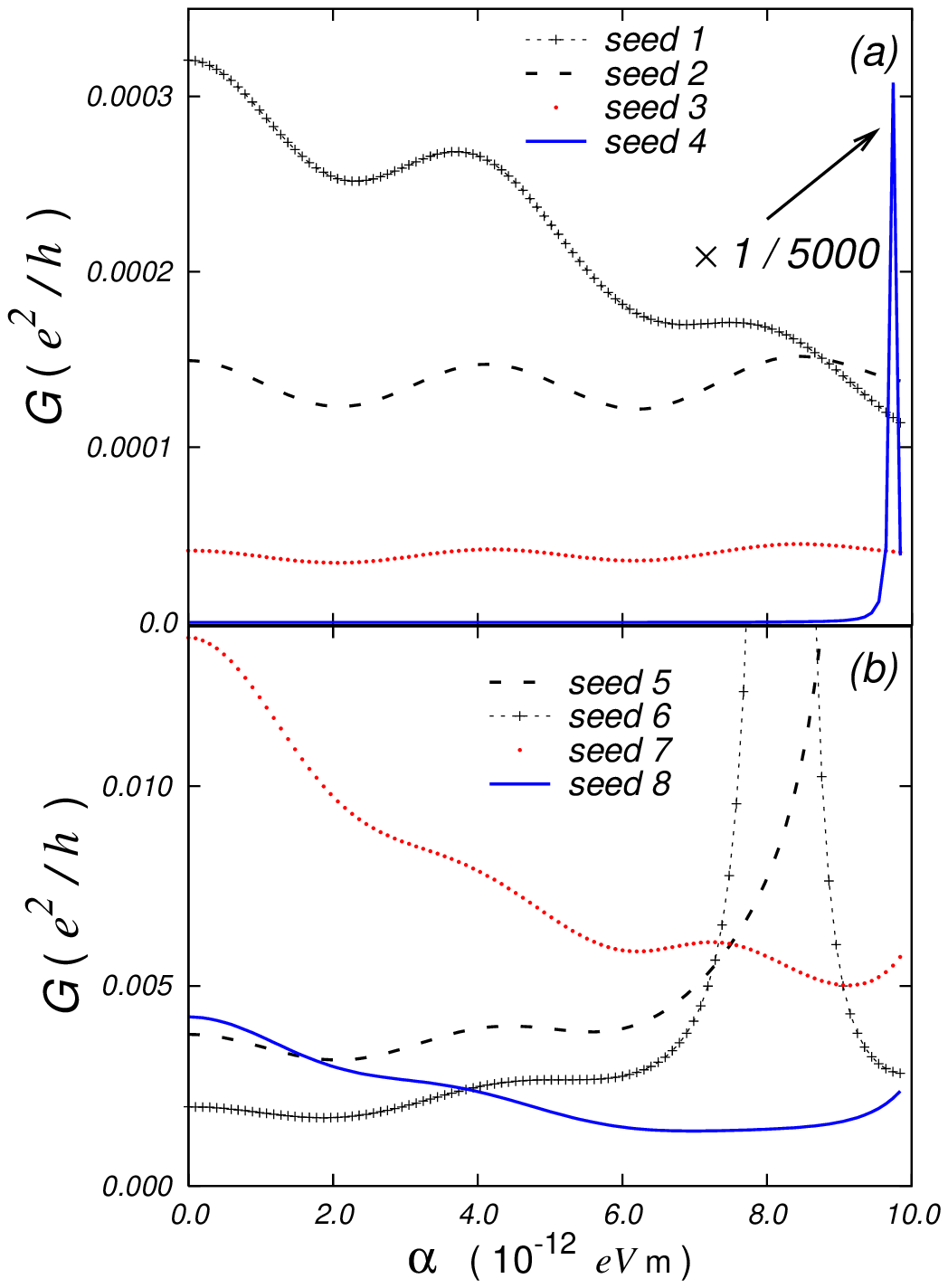}}
\caption{(Color online) The conductance $G$ for several impurity-configurations
in the 2DEG
as a function of the RSO coupling parameter
$\alpha$ for $N_{\rm ch}=3$ in (a) and $N_{\rm ch}=10$ in (b).
The seed numbers represent different impurity configurations,
all of which have the same mean free path $l=L/3$.
The other parameters are the same as in Fig.~\ref{non1}(a) and (b), respectively.
The length $L$ of the 2DEG is $1.44\,\mu$m.
}
\label{impurity}
\end{figure}

%
To find the origin of the large peaks, their dependence on the strengths of the hopping
parameters $t_{s_x}$ is examined.
As demonstrated in Fig.~\ref{non3}, the peak widths for the large
hopping parameters ($t_{s_x=1}\!=\!3\!\times\!0.1t$, $t_{s_x=-1}\!=\!3\!\times\!0.073t$)
are larger than those for the small hopping parameters
($t_{s_x=1}\!=\!0.1t$, $t_{s_x=-1}\!=\!0.073t$).
This difference in the peak widths implies that
the large peaks are resonances due to the electron confinement effects of the insulators.
Then at the energies where the resonances appear,
the phase acquired by electrons during their motion
from the injector to the collector should be an
integer multiple of $\pi$.
Interestingly this condition is the same as the condition for
the occurrence of the nested peak structure in Sec.~\ref{CoherentSFET}.
The comparison between the inset in Fig.~\ref{non3}(a) for the large
resonance peaks and Fig.~\ref{peak05}(b) for the nested peaks indeed verifies
this relation between the two phenomena.
Recalling that the inter-channel coupling is weak for $N_{\rm ch}\lesssim10$
and $\alpha\lesssim10\times10^{-12}$ eVm,
the correlation between the number of resonance peaks and
$N_{\rm ch}$ can also
be explained easily since each channel separately gives rise to resonances.

Lastly we examine effects of the nonideal spin injector/collector on a shorter SFET.
Figure~\ref{non4} shows the conductance evolution
for a SFET with relatively short $L=0.1\times1.44\,\mu$m in a much wider
window of $\alpha$, $(0\sim100)\times10^{-12}$ eVm.
Again large resonance peaks appear,
distorting the spin-orbit-coupling-induced modulation
of the conductance considerably.
We remark that the decrease of $L$ clearly enhances the spacing of the resonances
both in the $G$ vs. $\alpha$ plot and in the $G$ vs. $E_F$ plot (not shown).
However considering that the decrease of $L$ also widens the minimum size
of the $\alpha$ ``window'' for the observation of the spin-orbit-coupling-induced
modulation of $G$ and that the change of $E_F$ simply
``parallel'' shifts the resonance positions in the $G$ vs. $\alpha$ plot,
reducing $L$ does not appear to significantly facilitate the observation
of the spin-orbit-coupling-induced modulation of $G$.
%
\subsection{Scattering effects}
\label{scattering}
%
Another practical problem ignored in the preceding sections is
the scattering effects.
Scattering by impurities, which are isotropic, short-ranged, and spin-independent,
can be modelled by using the TB Hamiltonian of the impurity potential given by
$H^{\rm TB}_{\rm impurity}=\sum_{s_z}\sum^{N_x}_{i=1}\sum^{N_y}_{j=1}U_{i,j}c_{i,j,s_z}^\dagger c_{i,j,s_z}$,
where $U_{i,j}$ is a random variable whose value is uniformly distributed in the range $[-W, W]$.
We investigate briefly the scattering effects in a weakly diffusive regime;
$W$ is chosen in such a way that the mean free path $l$ is $L/3$.
The nonideal spin injector/collector is again assumed and $L=1.44\mu m$, $E_F=0.398t$ are used.
Figure~\ref{impurity} shows $G$ as a function of $\alpha$ for various realizations of
the random potential $U_{i,j}$.
For all realizations, $l$ is the same.
Note that for certain realizations, the spin-orbit-coupling-induced modulation of $G$ is
clearly visible while for other realizations, the modulations is strongly distorted by the
resonance peaks.
This difference is due to the fact that the positions of the resonances depend on
details of the spatial profile of the random potential $U_{i,j}$.
A similar result is obtained for a nonballistic single-channel SFET~\cite{Cahay03Cond}.
The results in Fig.~\ref{impurity} also imply that the conductance of a nonballistic multichannel SFET
equipped with the nonideal spin injector/collector may show considerable
sample-to-sample fluctuations.
Recalling that there are more resonance peaks for larger $N_{\rm ch}$,
the sample-to-sample fluctuations are expected to be more
significant for larger $N_{\rm ch}$.
More systematic studies about nonballistic multichannel SFETs
with the nonideal injector/collector are necessary.
%
%
\section{conclusion}
\label{Summary}
The issue of multichannel effects were addressed.
Via the numerical conductance calculation using the tight-binding approximation,
it has been demonstrated that the conductance modulation becomes less ideal
with the increase of the channel number $N_{\rm ch}$.
For the ideal spin injector/collector,
the conductance modulation ratio $G_{\rm max}/G_{\rm min}$
as a function of $N_{\rm ch}$ (from a few to 90) has been examined
for the RSO coupling parameter $\alpha\sim 5\times 10^{-12}$ eVm
and for the larger $\alpha\sim 50\times 10^{-12}$ eVm.
It has been found that
the decay of the modulation ratio with $N_{\rm ch}$
is considerably faster for the larger $\alpha$.
Through the study of the energy dispersion relations and the spin configurations of eigenstates,
it has been revealed that with the increase of $N_{\rm ch}$ and $\alpha$,
the spin configuration deviates further from the ideal 1D configuration
due to the formation of the phase-slip-like structure and the gradual drift of the spin angle,
and that the energy dispersion deviates further from the ideal quadratic dispersion
since the avoided crossing distorts the dispersion in a wider range of the longitudinal wavevectors.
It turns out that those deviations are rather small when $2m^*\alpha w/\hbar^2 \ll 2\pi$
while they become significant when $2m^*\alpha w/\hbar^2 \gtrsim 2\pi$.
Thus this result provides insights to the regime $2m^*\alpha w /\hbar^2 \gtrsim 2\pi$,
where the inter-channel coupling is expected to be significant~\cite{Datta90APL,Mireles01PRB}.
Effects of the electronic coherence in the multichannel regime were also examined and
we have found the nested peak structure,
which was attributed to the coherent Fabry-Perot-like interference in the multichannel regime.
We have also verified that the magnetic field-induced peak splitting,
reported previously~\cite{Lee05PRB} for a single-channel SFET, persists in the multichannel regime.
For the nonideal spin injector/collector that consists of a conventional metallic
ferromagnet-thin insulator-2DEG hybrid structure,
it has been found that large resonance peaks due to the electron confinement by the insulators
may strongly distort the spin-orbit-coupling-induced modulation of the conductance.
When the 2DEG becomes weakly diffusive, the conductance modulation signal shows
considerable sample-to-sample fluctuations.

\begin{acknowledgments}
We acknowledge H.-S. Sim, Hyowon Park, Woojoo Sim, and J.-H. Kim for their help on
numerical calculations.
This work was supported by the SRC/ERC program (Grant No.
R11-2000-071) and the Basis Research Program (Grant No.
R01-2005-000-10352-0) of MOST/KOSEF,
by the POSTECH Core Research Program,
and by the Korea Research Foundation Grant
(Grant No. KRF-2005-070-C00055 and BK21 program) funded by the
Korean Government (MOEHRD).
\end{acknowledgments}

\appendix
\section{Ideal injector and collector}
\label{IdealInjCol}

As a test of  the ideal spin injector,
we calculate the electron transmission probability $T_{\rm interface}$
through the injector-2DEG interface
for an electron incident from the injector side
with the incident momentum ${\bf k}=(k_x,k_y)$
and its spin pointing along the favored direction, $x$-axis.
The model Hamiltonian of the injector-2DEG hybrid system reads
$H_{\rm hyb}=\vartheta(x)H_{\rm 2D}+\vartheta(-x)H_{\rm inj}$,
where $\vartheta(x)$ is the Heaviside step function.
Since we are mainly interested in interface effects on the spin injection,
the confinement potential $V_{\rm c}$ in $H_{\rm 2D}$ and $H_{\rm inj}$
is ignored below.
For $\alpha=0$, $T_{\rm interface}$ is $1$
since $H_{\rm 2D}$ and $H_{\rm inj}$ are  identical
for the electron with its spin pointing along the $x$-axis.
For nonzero $\alpha$, $T_{\rm interface}$ will deviate from one.
%
\begin{figure}[b!]
\centerline{\includegraphics[height=9cm, width=4cm,
angle=-90]{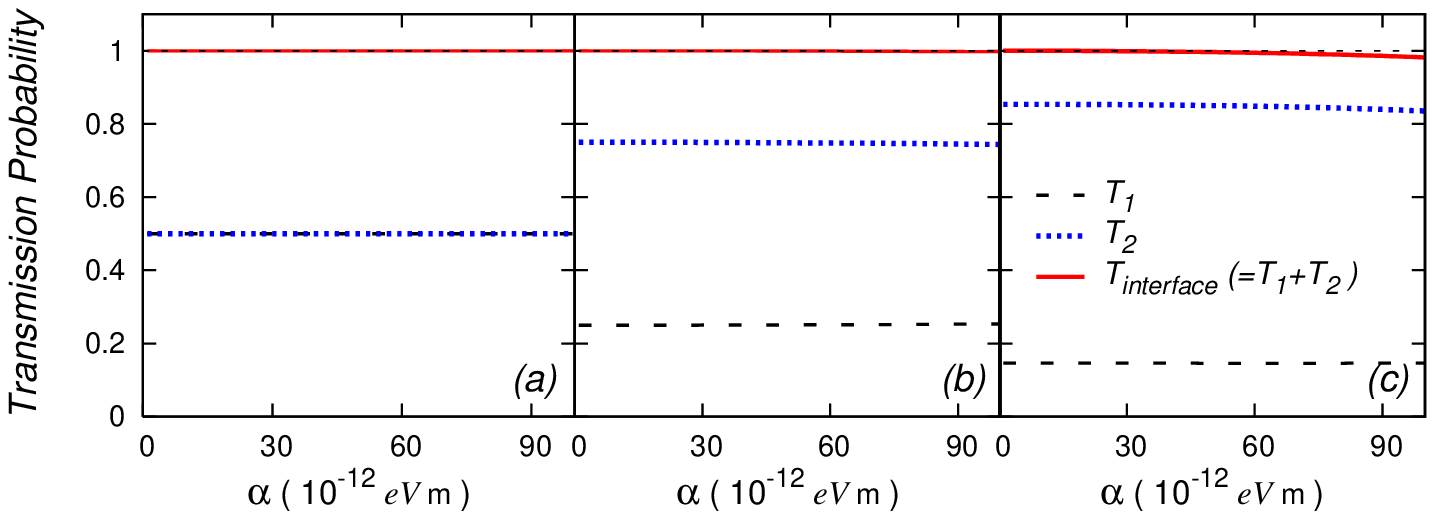}} \caption{(Color online) Transmission
probability $T_{\rm interface}$ through the injector-2DEG
interface as a function of the RSO coupling parameter $\alpha$.
The electron is injected from the injector side with its spin
pointing along the $(+x)$-direction and with the incident angle
$\theta$ (0 for normal incidence). (a) $\theta=0$, (b)
$\theta=\pi/6$ and (c) $\theta=\pi/4$. $T_1$ and $T_2$
($T_1+T_2=T_{\rm interface}$) are the transmission probabilities
into the two involved eigenstates in the 2DEG. } \label{interface}
\end{figure}
%
The evaluation of $T_{\rm interface}$
is straightforward. Recalling that the eigenspin direction in the
2DEG differs from $(+x)$-direction and that the injected electron
wave is thus a superposition of two plane waves with different
spin directions, one needs to match properly the superposition
with the electron state in the injector. The result is shown in
Fig.~\ref{interface} for the three incident angles $\theta=0$,
$\pi/6$, and $\pi/4$, where $\theta \equiv \sin^{-1}(k_y/|{\bf
k}|)$. Note that regardless of $\theta$, $T_{\rm interface}$
remains close to 1 up to considerably large $\alpha$,
$100\times10^{-12}$ eVm. Note also that $T_1$ and $T_2$
($T_1+T_2=T_{\rm interface}$), transmission probabilities into
each of the two plane wave states in the 2DEG, remain close to
their values $\cos^2(\theta/2+\pi/4)$ and $\sin^2(\theta/2+\pi/4)$
in the limit $\alpha\rightarrow 0$, indicating that the spin of
the injected electron points along the $x$-direction right after
the injection. Figure~\ref{interface} thus verifies the ideal spin
injector. The ideal spin collector can be verified in a similar
way.

\section{Symmetry analysis}
\label{Symmetry}
Symmetries often facilitate the analysis of a system and provide useful information~\cite{Zhai05PRL}.
Since $H_{\rm 2D}$ [Eq.~(\ref{eq:h1})] is invariant under the translation
along the $x$-axis,
an energy eigenfunction can be written as
$
\Psi(x,y)
=e^{ik_x x}[\psi_{\uparrow}(y)\chi_{\uparrow}
+\psi_{\downarrow}(y)\chi_{\downarrow}],
$
where $\psi_{\uparrow,\downarrow}(y)$ is the spin component
along the $(\pm z)$-direction,
$\chi_\uparrow \doteq\!{\tiny\left(\begin{array}{c}1\\0\\ \end{array}\right)}$
Pauli spin-up state and
$\chi_\downarrow \doteq\!{\tiny\left(\begin{array}{c}0\\1\\ \end{array}\right)}$
spin-down state.

$H_{\rm 2D}$ also commutes with two other symmetry operators $\Theta$ and $\Pi_x{\cal D}_{\rm spin}(\hat{\bf x},\pi)$,
where $\Theta$ is the time-reversal operator,
$\Pi_x$ is the mirror reflection operator with respect to the $yz$-plane in the orbital space,
and ${\cal D}_{\rm spin}(\hat{\bf x},\pi)$ is the rotation operator with respect to the $x$-axis by the angle $\pi$
in the spin space.
To examine implications of these symmetry operators, it is convenient to use
the representation ${\cal D}_{\rm spin}(\hat{\bf x},\pi)=\exp(-i\pi S_x/\hbar)\doteq -i\sigma_x$,
and $\Theta \doteq \sigma_y K$,
where $K$ is the complex-conjugation operator
in the real space representation with the $z$-axis as the spin quantization axis.
Application of $\Theta\Pi_x {\cal D}_{\rm spin}(\hat{\bf x},\pi)$ on $\Psi(x,y)$ results in
$e^{ik_x x}[\psi^*_{\uparrow}(y)\chi_{\uparrow}
-\psi^*_{\downarrow}(y)\chi_{\downarrow}]$.
Since $[H_{\rm 2D},\Theta\Pi_x {\cal D}_{\rm spin}(\hat{\bf x},\pi)]=0$,
the resulting state should also be an energy eigenfunction
with the energy eigenvalue same as that for $\Psi(x,y)$.
Then by comparing $\Psi(x,y)$ with $\Theta \Pi_x{\cal D}_{\rm spin}(\hat{\bf x},\pi)\Psi(x,y)$,
one finds that without loss of generality, $\psi_{\uparrow}(y)$ may be chosen to be real
and $\psi_{\downarrow}(y)$ may be chosen to be imaginary.
Recalling the spinor representation
$\left( {\tiny \begin{array}{c} \cos\theta/2 \\ e^{i\phi}\sin\theta/2 \end{array} } \right)$
for the spin with the polar angle $\theta$
and the azimuthal angle $\phi$,
one finds that $\phi$ for the energy eigenfunctions is $\pm \pi/2$ and
that the spin direction of the energy eigenfunctions always lies within the $yz$-plane.

When the confining potential is symmetric $V_{\rm c}(y)=V_{\rm c}(w-y)$,
which is the case for the hard wall confinement potential,
$H_{\rm 2D}$ allows another symmetry operation $i\Pi_y {\cal D}_{\rm spin}(\hat{\bf y},\pi)$,
where $\Pi_y$ is the mirror reflection operator with respect to the $y=w/2$ plane in the orbital space
and ${\cal D}_{\rm spin}(\hat{\bf y},\pi)$ is the rotation operator
with respect to the $y$-axis by the angle $\pi$ in the spin space.
Since $[i\Pi_y{\cal D}_{\rm spin}(\hat{\bf y},\pi)]^2=I$,
possible eigenvalues of $i\Pi_y {\cal D}_{\rm spin}(\hat{\bf y},\pi)$
are $+1$ and $-1$, and
the energy eigenfunctions can be classified into even and odd parity states accordingly.
For notational convenience, we choose $\phi=+\pi/2$ for even parity states
and $\phi=-\pi/2$ for odd parity states.
At the special point $y=w/2$, $\Pi_y$ plays no role
and the parity becomes identical to the eigenvalue
of $i{\cal D}_{\rm spin}(\hat{\bf y},\pi)=\sigma_y$.
Thus at $y=w/2$, the spin of the even parity states points towards $(+y)$-direction
($\theta=\pi/2$, $\phi=+\pi/2$)
and the spin of odd parity states points towards $(-y)$-direction
($\theta=\pi/2$, $\phi=-\pi/2$).
For $y\neq w/2$, the spin in general deviates from $(\pm y)$-direction.
But due to the parity constraint, the spin directions at $y$ and $w-y$ are correlated;
$\theta(y)+\theta(w-y)=\pi$.

\section{Hard wall confinement : Exact results}
\label{ExactDispersion}
An exact energy eigenstate
$\Psi(x,y)=e^{ik_x x}
[\psi_{\uparrow}(y)\chi_{\uparrow}
+\psi_{\downarrow}(y)\chi_{\downarrow}]$
of $H_{\rm 2D}$ [Eq.~(\ref{eq:h1})] with energy $E$ can be expressed as a linear superposition of
plane waves as follows,
\begin{equation}
\label{eq:eigenft}
\Psi(x,y)=\sum_{j=1}^4 c_j \Psi_{{\bf k}_j}(x,y),
\end{equation}
where the plane wave $\Psi_{{\bf k}_j}(x,y)= e^{i{\bf k}_j\cdot {\bf r}}u_j({\bf k}_j)$
is an energy eigenfunction of $H_{\rm 2D}$ in the absence of
the confinement potential [$V_c(y)=0$],
${\bf r}=(x,y)$, ${\bf k}_j=(k_x,k_{y,j})$,
and $k_{y,j}$'s are four solutions of
$E=\hbar^2 (k_x^2+k_{y,j}^2)/2m^*
+(-1)^j|\alpha|\, (k_x^2+k_{y,j}^2)^{1/2}$ (see Fig.~\ref{2D}).
When all four solutions are real (denoted by the blue solid line in Fig.~\ref{2D}),
one may take
$k_{y,1}=-k_{y,3}$ and $k_{y,2}=-k_{y,4}$.
But in certain situations (denoted by the red dashed line in Fig.~\ref{2D}),
only two solutions are real and the other two remaining solutions are pure imaginary.
We remark that in order to obtain a complete energy spectrum,
one should consider such situations as well.
In such situations, one may take $k_{y,1}=-k_{y,3}$ for the two real solutions
and $k_{y,2}^*=-k_{y,4}$ for the other two purely imaginary solutions.
The procedure described below applies to both situations (blue solid line and red dashed line) denoted in Fig.~\ref{2D}.
For the spinors $u_1({\bf k}_1)$ and $u_2({\bf k}_2)$, we choose the representation,
\begin{equation}
\label{eq:spinor02-1}
u_1({\bf k}_1)={1\over\sqrt{2}}
\left(\begin{array}{c}1 \\ i{\alpha \over |\alpha|}{k_x+ik_{y,1} \over (k_x^2+k_{y,1}^2)^{1/2}}
\end{array}\right),
\end{equation}
and
\begin{equation}
\label{eq:spinor03}
u_2({\bf k}_{2})={1\over\sqrt{2}}
\left(\begin{array}{c} 1 \\ -i {\alpha \over |\alpha|}{k_x+ik_{y,2} \over (k_x^2+k_{y,2}^2)^{1/2}}
\end{array}\right).
\end{equation}
For the spinors $u_3({\bf k}_3)$ and $u_4({\bf k}_4)$, we choose their representation
in such a way that the plane waves $\Psi_{{\bf k}_j}(x,y)$'s satisfy the relations,
$\Psi_{{\bf k}_3}(x,y)=i\Pi_y{\cal D}_{\rm spin}(\hat{\bf y},\pi)\Psi_{{\bf k}_1}(x,y)$
and $\Psi_{{\bf k}_4}(x,y)=i\Pi_y{\cal D}_{\rm spin}(\hat{\bf y},\pi)\Psi_{{\bf k}_2}(x,y)$,
which is possible since $i\Pi_y{\cal D}_{\rm spin}(\hat{\bf y},\pi)$ is a symmetry operation
of the system.
%
%

The four coefficients $c_j$'s in Eq.~(\ref{eq:eigenft}) are fixed by the hard-wall boundary conditions,
$\Psi(x,y=0)=\Psi(x,y=w)=0$,
which lead to a secular equation for a $4\times 4$ matrix.
Solving this equation is facilitated by the symmetry operation $i\Pi_y{\cal D}_{\rm spin}(\hat{\bf y},\pi)$.
Since $i\Pi_y{\cal D}_{\rm spin}(\hat{\bf y},\pi)$ has only two eigenvalues, $+1$ and $-1$,
the eigenstates can be classified into even and odd parity states.
For an even parity state, one may choose $c_3=c_1$ and $c_4=c_2$,
and for an odd parity state, one may choose $c_3=-c_1$ and $c_4=-c_2$.
Thus the secular equation for the $4\times 4$ matrix is block-diagonalized
into two $2\times 2$ matrices, one for even parity states
and the other for odd parity states.
By solving the two $2\times 2$ matrices separately, one obtains the eigenenergy $E$ and
the corresponding eigenstate $\Psi(x,y)$ for even and odd parity states.
%
Figure~\ref{energy}  shows the energy dispersion
for various combinations of $\alpha$ and $w$.
Figure~\ref{SpinTexture} shows the spin profiles of selected eigenstates.

\section{Perturbation theory}
\label{PerturbationTheory}
Although the exact eigen wavefunctions and energy-dispersion relations are readily available,
perturbation theory calculations are still useful
in order to gain insights into
the $\alpha$-dependence of eigen transport channels.
Below we treat the RSO coupling as a perturbation and
examine its effects on eigen transport channels.
Since the longitudinal momentum $k_x$ is an exact quantum number of $H_{\rm 2D}$ in Eq.~(\ref{eq:h1}),
one may work with the reduced Hamiltonian $H_{\rm 2D}(k_x)=H_0+H_R$
that acts on the transverse part $\psi(y)$ of the original wavefunction $\Psi(x,y)=e^{ik_x x}\psi(y)$.
Here the unperturbed Hamiltonian $H_0$ and the perturbation $H_R$ are given by
\begin{equation}
\label{eq:H02}
H_0\!\equiv \!\left[{\hbar^2k^2_x\over2m^*}\!+\!{p^2_y\over2m^*}\!+\!V_{\rm c}(y)\right]\!I,
\phantom{1}
%
%
H_R\!\equiv \!\alpha\left(\sigma_x {p_y\over \hbar}-\sigma_y k_x\right),
\end{equation}
where $I$ is the identity operator acting on spin.

In the zeroth order in $\alpha$, each energy eigenvalue of $H_{\rm 2D}(k_x)$ is doubly degenerate
with
$
E^{(0)}_{n,1}=E^{(0)}_{n,2}=\hbar^2k^2_x/2m^*+\mathcal{E}_n,
$
where $\mathcal{E}_n\equiv \hbar^2 k_{n,y}^2/2m^*$
is the energy of the $n$-th ($n$=0,1,2,$\cdots$) excitation mode in the transverse motion
and $k_{n,y}\equiv (n+1)\pi/w$.
The corresponding eigenfunctions
are
$
\psi^{(0)}_{n,1}(y)\!=\!\phi_n(y)\cdot
(\chi_\uparrow+i\chi_\downarrow)/\sqrt{2}
$
and
$
\psi^{(0)}_{n,2}(y)\!=\!\phi_n(y)\cdot
(\chi_\uparrow-i\chi_\downarrow)/\sqrt{2}
$
respectively, where $\phi_n(y)\equiv \sqrt{2/w}\, \sin k_{n,y}y$,
and
$\chi_\uparrow=\left( {\tiny \begin{array}{c} 1 \\ 0 \end{array} } \right)$,
$\chi_\downarrow=\left( {\tiny \begin{array}{c} 0 \\ 1 \end{array} } \right)$
are the spinors pointing along the $(\pm z)$-directions.

Corrections to the eigenenergies can be obtained easily.
The first order corrections to $E^{(0)}_{n,i}$ ($i=$1,2) are given by
the representation of $H_R$ within the two-dimensional
degenerate subspace:
\begin{equation}
\label{eq:matHR}
\left(\begin{array}{cc}
\langle\psi^{(0)}_{n,1}|H_R|\psi^{(0)}_{n,1}\rangle &
\langle\psi^{(0)}_{n,1}|H_R|\psi^{(0)}_{n,2}\rangle \\
\langle\psi^{(0)}_{n,2}|H_R|\psi^{(0)}_{n,1}\rangle &
\langle\psi^{(0)}_{n,2}|H_R|\psi^{(0)}_{n,2}\rangle \\
\end{array}\right)
\!=\!
\left(\begin{array}{cc}
-\alpha k_x & 0 \\
0 & \alpha k_x \\
\end{array}\right).
\end{equation}
%
Thus one obtains the first order energy correction
$E^{(1)}_{n,i}=(-1)^i\alpha k_x$.
Note that since the $2\!\times\!2$ matrix of
Eq.~(\ref{eq:matHR}) is diagonal,
$\psi_{n,1}^{(0)}$ and $\psi_{n,2}^{(0)}$ are
the proper zeroth order eigenfunctions in view of the degenerate perturbation theory.
The second order energy corrections are given by
\begin{equation}
\label{eq:secondE01}
E_{n,i}^{(2)}=\langle\psi_{n,i}^{(0)}|H_R{\mathcal{P}_n\over E_{n,i}^{(0)}-H_0}H_R|
\psi_{n,i}^{(0)}\rangle,
\end{equation}
where the projection operator $\mathcal{P}_n$ is defined by
$
\mathcal{P}_n\equiv 1
-\sum_{i=1}^2|\psi_{n,i}^{(0)}\rangle\langle\psi_{n,i}^{(0)}|
$.
By using the identity
$\mathcal{P}_n\sigma_y|\psi_{n,i}^{(0)}\rangle=0$
and the relation $[H_0, \sigma_x]=0$,
Eq.~(\ref{eq:secondE01}) can be written as
\begin{equation}
\label{eq:secondE02}
E_{n,i}^{(2)}={\alpha^2 \over \hbar^2}\sum_{l\neq n}
{1\over \mathcal{E}_n-\mathcal{E}_l}
\left|\int dy\phi_l^*(y)p_y
\phi_n(y)\right|^2,
\end{equation}
which reduces further to
\begin{equation}
\label{eq:secondE03}
E_{n,i}^{(2)}=-{m^*\alpha^2\over2\hbar^2},
\end{equation}
with the help of the identities $p_y=(m^*/i\hbar)[y,H_0]$
and $\sum_l \phi^*_l(y')\phi_l(y)=\delta(y-y')$.
%
%
Thus up to the second order in $\alpha$,
the energy dispersion relation is given by
\begin{equation}
\label{eq:eigenenergy-2}
E_{n,i}={\hbar^2k_x^2\over 2m^*}+\mathcal{E}_n+(-1)^{i}\alpha k_x
-{m^*\alpha^2\over2\hbar^2}\, .
\end{equation}
The result is shown in Fig.~\ref{energy}.
Here we remark that although Eq.~(\ref{eq:eigenenergy-2}) is derived for
the hard-wall confinement potential, the first- and second-order corrections to the eigenenergy
are independent of $V_{\rm c}(y)$; its change alters only the zeroth order contribution $\mathcal{E}_n$.
A similar result for a harmonic confinement has been reported previously~\cite{Serra05PRB}.

Next we examine corrections to energy eigenstates.
According to the degenerate perturbation theory, the first order
corrections
$|\psi_{n,i}^{(1)}\rangle=\mathcal{P}_n|\psi_{n,i}^{(1)}\rangle
+\mathcal{Q}_n|\psi_{n,i}^{(1)}\rangle$, where
$\mathcal{Q}_n \equiv \sum_{i=1}^2|\psi_{n,i}^{(0)}\rangle\langle\psi_{n,i}^{(0)}|$
and $\mathcal{P}_n+\mathcal{Q}_n=1$, are given by
\begin{equation}
\label{eq:Pn01}
\mathcal{P}_n|\psi_{n,i}^{(1)}\rangle\!=\!
\sum_{m\neq n}\!\sum_{j=1}^2
|\psi_{m,j}^{(0)}\rangle{1\over E_{n,i}^{(0)}\!-\!E_{m,j}^{(0)}}
\langle\psi_{m,j}^{(0)}|H_R|\psi_{n,i}^{(0)}\rangle,
\end{equation}
and
\begin{equation}
\label{eq:Qn01}
\mathcal{Q}_n|\psi_{n,i}^{(1)}\rangle\!=\!|\psi_{n,\bar{i}}^{(0)}\rangle\!
{1\over E_{n,i}^{(1)}\!-\!E_{n,\bar{i}}^{(1)}}\!
\langle\psi_{n,\bar{i}}^{(0)}\!|H_R\!{P_n\over E_n^{(0)}\!-\!H_0}\!H_R|
\psi_{n,i}^{(0)}\rangle,
\end{equation}
where $\bar{i}=2$ for $i=1$ and $\bar{i}=1$ for $i=2$.
Both Eqs.~(\ref{eq:Pn01}) and (\ref{eq:Qn01}) can be considerably simplified.
With the help of the identities
$\mathcal{P}_n\sigma_y|\psi_{n,i}^{(0)}\rangle=0$,
$p_y=(m^*/i\hbar)[y,H_0]$,
and $\sum_l \phi^*_l(y')\phi_l(y)=\delta(y-y')$,
one finds
\begin{equation}
\label{eq:Pn02}
\mathcal{P}_n|\psi_{n,i}^{(1)}\rangle
=(-1)^{i+1}{m^*\alpha\over \hbar^2} (y-\langle y \rangle_n)|\psi_{n,\bar{i}}^{(0)}\rangle,
\end{equation}
\begin{equation}
\label{eq:Qn02}
\mathcal{Q}_n|\psi_{n,i}^{(1)}\rangle=0,
\end{equation}
where $\langle y\rangle_n\equiv\int dy \phi_n^*(y) y \phi_n(y)=w/2$.
Then up to the first order in $\alpha$,
the eigen wavefunctions are given by
\begin{equation}
\label{eq:state01}
\psi_{n,i}(y) =
\psi_{n,i}^{(0)}(y)
+(-1)^{i+1}{m^*\alpha(y-\langle y\rangle_n)\over\hbar^2}
\, \psi_{n,\bar{i}}^{(0)}(y),
\end{equation}
%
%
We remark that although Eq.~(\ref{eq:state01}) is derived for the hard wall confinement,
it holds for other confinement potentials $V_{\rm c}(y)$ as well
when $\langle y \rangle_n$ is evaluated for the new $V_{\rm c}(y)$.
The second order correction
$|\psi_{n,i}^{(2)}\rangle=\mathcal{P}_n|\psi_{n,i}^{(2)}\rangle
+\mathcal{Q}_n|\psi_{n,i}^{(2)}\rangle$ is given by
\begin{eqnarray}
\mathcal{P}_n|\psi_{n,i}^{(2)}\rangle=&{\mathcal{P}_n \over E_{n,i}^{(0)}-H_0}
(H_R-E_{n,i}^{(1)})\mathcal{Q}_n |\psi_{n,i}^{(1)}\rangle \nonumber\\
&+\left[ {\mathcal{P}_n \over E_{n,i}^{(0)}-H_0}(H_R-E_{n,i}^{(1)}) \right]^2
|\psi_{n,i}^{(0)}\rangle,
\end{eqnarray}
\begin{eqnarray}
\mathcal{Q}_n|\psi_{n,i}^{(2)}\rangle=&|\psi_{n,\bar{i}}^{(2)}\rangle
{1\over E_{n,i}^{(1)}-E_{n,\bar{i}}^{(1)}}\left(
\langle \psi_{n,\bar{i}}^{(0)}|H_R \mathcal{P}_n|\psi_{n,i}^{(2)}\rangle \right.\nonumber\\
&\left.-{E_{n,i}^{(2)} \over E_{n,i}^{(1)}-E_{n,\bar{i}}^{(1)}}
\langle \psi_{n,\bar{i}}^{(0)}|H_R \mathcal{P}_n|\psi_{n,i}^{(1)}\rangle
\right).
\end{eqnarray}
With the help of similar identities, one finds
\begin{eqnarray}
&\mathcal{P}_n|\psi_{n,i}^{(2)}\rangle
&=\!\!\left({m^*\alpha w \over \hbar^2} \right)^2\!\!
\left[2\!\!\sum_{n'\neq n}{\hbar^2/m^* w^2 \over E_{n,i}^{(0)}-E_{n',\bar{i}}^{(0)} }
|\psi_{n',\bar{i}}^{(0)}\rangle
(k_x \langle y \rangle_{n',n}) \right.
\label{P2} \nonumber \\
&-&\!\!\!\!\!\left.\sum_{n',n''\neq n}\!\!\!{E_{n',i}^{(0)}-E_{n'',i}^{(0)} \over E_{n,i}^{(0)}-E_{n'',i}^{(0)} }
|\psi_{n'',i}^{(0)}\rangle
{\langle y \rangle_{n'',n'} \over w}{\langle y \rangle_{n',n} \over w}
\right], \\
&\mathcal{Q}_n|\psi_{n,i}^{(2)}\rangle&= 0,
\label{Q2}
\end{eqnarray}
%
where $\langle y \rangle_{n,n'}\equiv \langle \phi_n|y|\phi_{n'}\rangle$.
We remark that these expressions for the second correction are again
independent of the choice of $V_{\rm c}(y)$.

These expressions contain the information on the local spin
direction.
Since the Hamiltonian $H_{\rm 2D}$ is invariant under the transformation
$\Theta \Pi_x {\cal D}_{\rm spin}(\hat{\bf x},\pi)$,
the local spin direction is always perpendicular to the $x$-axis as demonstrated in Appendix~\ref{Symmetry}
and it can be parameterized by a single angle $\theta_{n,i}(y)$
within the $yz$ plane;
$
\psi_{n,i}(y)=
|\psi_{n,i}(y)|\cdot
\left(\cos{\theta_{n,i}(y)\over 2}\chi_{\uparrow}
+i\sin{\theta_{n,i}(y)\over 2}\chi_{\downarrow}\right).
$
The local spin direction is pointing $-y$, $+z$, $+y$,  and $-z$ axis
when $\theta_{n,i}$ is $-\pi/2$, $0$, $\pi/2$, and $\pm\pi$, respectively.
By using the perturbatively obtained wavefunctions,
the angle $\theta_{n,i}(y)$ can be expanded as a power series of $\alpha$,
$\theta_{n,i}=\theta^{(0)}_{n,i}+\theta^{(1)}_{n,i}+\theta^{(2)}_{n,i}+\cdots$,
where the zeroth and the first order contributions are
\begin{equation}
\label{eq:theta01}
\theta_{n,1}^{(0)}(y)\!=\!{\pi\over2},\phantom{2}
\theta_{n,1}^{(1)}(y)\!=\!-{2m^*\alpha\over\hbar^2} (y-\langle
y\rangle_n),
\end{equation}
and
\begin{equation}
\label{eq:theta02}
\theta_{n,2}^{(0)}(y)\!=\!-{\pi\over2},\phantom{2}
\theta_{n,2}^{(1)}(y)\!=\!-{2m^*\alpha\over\hbar^2} (y-\langle
y\rangle_n).
\end{equation}
A similar result has been reported previously~\cite{Hausler01PRB}.
Note that the first order contributions change linearly with $y$,
explaining the gradual drift in Fig.~\ref{SpinTexture}.
Note also that despite the $y$-dependence of $\theta_{n,i}$
the local spin directions of $\psi_{n,1}(y)$ and $\psi_{n,2}(y)$
remain antiparallel to each other [$\psi_{n,1}^\dagger(y)\psi_{n,2}(y)=0$]
at every position $y$ up to the first order in $\alpha$
since $|\theta_{n,1}(y)-\theta_{n,2}(y)|=\pi$ up to the first order in $\alpha$.
The second order contribution $\theta_{n,i}^{(2)}(y)$ can be also obtained (expressions not shown)
from the perturbative expressions for $\psi_{n,i}(y)$
and it turns out that
when $\theta_{n,i}^{(2)}(y)$'s are taken into account,
the local spin directions of $\psi_{n,1}(y)$ and $\psi_{n,2}(y)$
are {\it not} antiparallel to each other any more.
Solid lines in Fig.~\ref{SpinTexture} show $\theta_{n,i}(y)$
calculated from the perturbative expressions for $\psi_{n,i}(y)$,
where it is clear that the deviation of $|\theta_{n,1}(y)-\theta_{n,2}(y)|$ from $\pi$
becomes more evident for larger $\alpha$.
The second order contribution $\theta_{n,i}^{(2)}(y)$ is responsible
for the phase-slip-like structure in Fig.~\ref{SpinTexture}.

\section{A nonideal spin injector}
\label{Injection}
\begin{figure}[b!]
\centerline{\includegraphics[width=5cm, height=8cm, angle=-90]{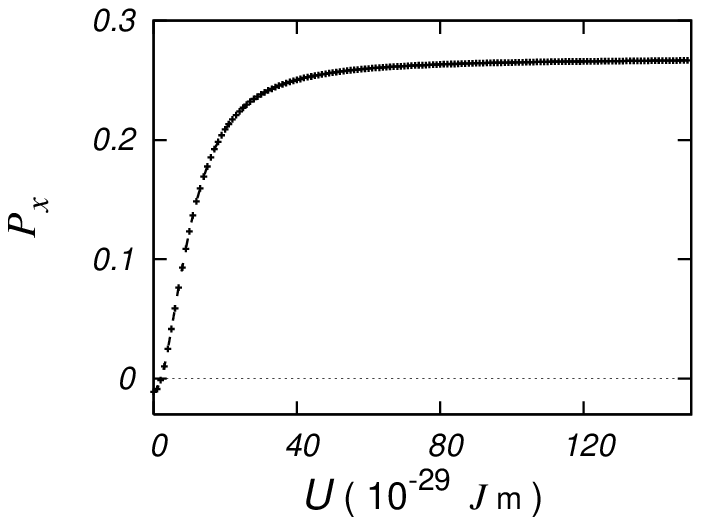}}
\caption{ Spin polarization of injected currents, $P_x$
whose spin is along the $(+x)$-direction,
as a function of the tunnel contact parameter $U$.}
\label{polarization}
\end{figure}

Here we consider the spin injection from
a nonideal injector, which consists of a conventional metallic
ferromagnet(F)-insulator(I)-2DEG heterostructure.
The differences of the effective masses,
the densities of states, and the Fermi wavelengths between the F
and the 2DEG are taken into account
and the electron transport within the F and the 2DEG is
assumed to be ballistic.
It has been reported~\cite{Teresa99PRL} that the detailed structure
of the spin dependent density of states within the insulator
can influence the spin polarization of the injected electron.
In this paper, however, we ignore such details for simplicity
and model the insulator
by a $\delta$-function potential $U\delta(x)$,
where $U$ denotes the tunnel contact parameter~\cite{Blonder82PRB,Hu01PRL}.
The model Hamiltonian of the F-I-2DEG hybrid system reads
$H'_{\rm hyb}=\vartheta(x)H_{\rm 2D}+\vartheta(-x)H_{\rm F}+U\delta(x)$,
where $H_{\rm F}$ has the form $H_{\rm F}={\mathbf p}^2/2m_{\rm electron}-h_0\sigma_x+V_0$.
Here, $h_0$ represents the exchange interaction of the Stoner model in the F,
and $V_0$ denotes the difference between the bottoms of the energy bands for the F and 2DEG.
Below we ignore  the confinement potential $V_c$ and focus on the case of
the normal incidence ($\theta=0$) from the F
since the Fermi wave vector in the F
is much larger than that of the 2DEG~\cite{Matsuyama02PRB}
and thus the normally incident electrons constitute a major portion of the injected electrons.
We choose the value of the Fermi energy and the exchange interaction $h_0$ of the F to be $3.500$ eV and
$1.750$ eV, respectively, and set the Fermi energy $E_F$ of the 2DEG to $E_F=0.398t=0.103$ eV.
Thus $V_0=(0.103-3.500)$eV.
Figure~\ref{polarization} shows the spin polarization $P_x$ of the injected currents
as a function of the tunnel contact parameter $U$,
where $P_x$ is a measure of the spin polarization along the $(+x)$-direction and is defined by
$(T^{s_x=1}_{\rm interface}-T^{s_x=-1}_{\rm interface})/(T^{s_x=1}_{\rm interface}+T^{s_x=-1}_{\rm interface})$.
Here $T^{s_x}_{\rm interface}$ represents the electron transmission probability through the F-I-2DEG
structure for an incident electron with the spin $s_x$.
Note that while $P_x$ is only a few percent in the absence of $U$,
it grows close to 30\%.
This demonstrates clearly that the spin-dependent tunneling over the insulator barrier is the main
source of the spin polarization of the injected current while the bulk properties of the F
are rather irrelevant factors for the spin polarization of the injected current.
This observation, which is in agreement with Refs.~\cite{Rashba00PRB,Hu01PRL},
justifies the simplified description in Sec.~\ref{nonidealInjCol} of the F-I-2DEG structure,
where the differences between the two bulks (F and 2DEG) are ignored
while the spin-dependent tunneling over the I is taken into account.




\begin{thebibliography}{99}
\bibitem{ThermalAverage}
For a single-channel SFET, it has been demonstrated~\cite{Lee05PRB}
that the $E_F$-dependence of the conductance profile is averaged out
at temperatures $k_B T$ sufficiently larger than $\pi \hbar v_F/L$,
where $v_F$ is the Fermi velocity,
and the high temperature expression is given by
$G(\alpha)=(2e^2/h)\cos^2(\phi_\alpha/2)/[1+\cos^2(\phi_\alpha/2)]$.
Then a trivial generalization to a multichannel SFET
is obtained by multiplying $N_{\rm ch}$ to the single-channel result.
\bibitem{Kn-comment} In the evaluation of $v_{F,n}$, smaller correction due to the RSO coupling is ignored.
In the presence of $B_\parallel$, the definition of $K_n$ is altered to
$K_n=[(k_x^{\rm right})_{n,i=1}+(k_x^{\rm right})_{n,i=2}
+|(k_x^{\rm left})_{n,i=1}|+|(k_x^{\rm left})_{n,i=2}|]/4$,
where $(k_x^{\rm right})_{n,i}$ $(>0)$ and $(k_x^{\rm left})_{n,i}$ $(<0)$
are the longitudinal wavevectors at $E_F$ for
the right and left moving electrons.
In the absence of $B_\parallel$, this definition reduces
to the definition right below Eq.~(\ref{eq:G1}).
\end{thebibliography}
\end{document}